\documentclass[aps,prb, preprint, citeautoscript, showpacs,floatfix,superscriptaddress]{revtex4-1}
\usepackage{bm}
\usepackage{amssymb}
\usepackage{graphicx}
\usepackage{amsmath}
\usepackage{textcomp}
\usepackage{colordvi}
\usepackage{multirow}
\usepackage{ulem}

\begin{document}
\title{Electric field induced injection and shift currents in zigzag graphene nanoribbons}
\author{Yadong Wei}
\affiliation{
 School of Physics, Harbin Institute of Technology, 92 Xidazhi Street, Nangang, Harbin, China
}
\affiliation{
 Changchun Institute of Optics, Fine Mechanics and Physics, Chinese Academy of Sciences, Changchun 130033, People’s Republic of China
}
\author{Weiqi Li}
\affiliation{
 School of Physics, Harbin Institute of Technology, 92 Xidazhi Street, Nangang, Harbin, China
}
\author{Yongyuan Jiang}
\affiliation{
  School of Physics, Harbin Institute of Technology, 92 Xidazhi Street, Nangang, Harbin, China
}
\affiliation{
  Key Lab of Micro-Optics and Photonic Technology of Heilongjiang Province, Harbin, China
}
\affiliation{
  Key Laboratory of Micro-Nano Optoelectronic Information System, Ministry of Industry and Information Technology, Harbin, China
}
\affiliation{
  Collaborative Innovation Center of Extreme Optics, Taiyuan 030006, Shanxi, China
}
\author{Jinluo Cheng}
 \email{jlcheng@ciomp.ac.cn}
 \affiliation{
 Changchun Institute of Optics, Fine Mechanics and Physics, Chinese Academy of Sciences, Changchun 130033, People’s Republic of China
}
 \affiliation{
 School of Physical Sciences, University of Chinese Academy of Sciences, Beijing 100049, People’s Republic of China
}

\begin{abstract}
We theoretically investigate the one-color injection currents and shift currents in zigzag graphene nanoribbons with applying a static electric field across the ribbon, which breaks the inversion symmetry to generate nonzero second order optical responses by dipole interaction. These two types of currents can be separately excited by specific light polarization, circularly polarized lights for injection currents and linearly polarized lights for shift currents. Based on a tight binding model formed by carbon 2p$_z$ orbitals, we numerically calculate the spectra of injection coefficients and shift conductivities, as well as their dependence on the static field strength and ribbon width.
The spectra show many peaks associated with the optical transition between different subbands, and the positions and amplitudes of these peaks can be effectively controlled by the static electric field. By constructing a simple two band model, the static electric fields are found to modify the edge states in a nonperturbative way, and their associated optical transitions dominate the current generation at low photon energies. For typical parameters, such as a static field 10$^6$ V/m and light intensity 0.1 GW/cm$^2$, the magnitude of the injection and shift currents for a ribbon with width 5 nm can be as large as the order of 1 $\mu$A. Our results provide a physical basis for realizing passive optoelectronic devices based on graphene nanoribbons.
\end{abstract}


\maketitle

\section{Introduction}

Graphene nanoribbon (GNR) is a narrow stripe of monolayer graphene
with width varying from a few nanometers to less than 50 nanometers, at
which it shows exciting physical properties in addition to graphene
due to the quantum confinement.\cite{PhysRevLett.97.216803}  Combining
with its compatibility with industry-standard lithographic processing
\cite{Nat.Mater._8_235_2009_Ritter,J.Phys.Chem.B_108_19912_2004_Berger}
and the increasingly mature fabrication procedure
\cite{ActaPhysicaSinica_68_168102_2019_Chen}, GNR is considered as a
potential material for applications in nanoelectronics and
optoelectronics. Many efforts have been devoted to understand its band
structures, transport properties, magnetism, chirality, optical
properties, and so on
\cite{grapheneoptoelectronic1,graphenetransport,NaturePhys_7_616_Chenggang_Tao,PhysRevB.59.8271}.

The widely studied GNRs include armchair GNRs (aGNRs) with edges orientated along the armchair directions and zigzag GNRs (zGNRs) with edges orientated along the zigzag directions. The band structures of GNRs have been calculated by different models, such as tight binding model\cite{PhysRevB.54.17954}, continuum model based on a $\bm k\cdot\bm p$ Hamiltonian\cite{PhysRevB.73.235411}, and first principle calculations\cite{PhysRevLett.97.216803}. 
 The simplest tight binding model shows that zGNR is always metallic with flat bands induced by edge states, and aGNR can be either semiconducting or metallic depending on its width \cite{PhysRevB.54.17954}. After considering the Coulomb interaction, DFT calculations show that all narrow GNRs have finite gaps, and zGNRs possess antiferromagnetic ground states \cite{PhysRevB.77.073412}. The band gap has a strong dependence on the edge orientation and ribbon width. In such tight binding model, both the eigenstates and selection rules of the optical transition can be analytically obtained, and many absorption peaks are induced by the optical transitions between different subbands\cite{PhysRevB.76.045418,PhysRevB.95.155438}. The linear optical response 
shows strong anisotropy along zigzag and armchair directions. With
applying an external static electric field across the ribbon, the gap
can be effectively tuned and becomes closed at an appropriate field
strength; and furthermore the optical properties are effectively modulated\cite{Carbon_44_508_Chang,JPCM_27_145305_Saroka}. Because of the insufficient Coulomb screening,
the excitonic effects are important for narrow ribbons\cite{PhysRevLett.101.186401,PhysRevB.88.165425,AZNRexciton,PhysRevB.99.165415}.

In addition to linear optical responses, the nonlinear optical
properties of GNR also attracted much attention. By tuning the doping level electrically, Cox
{\it et al.} studied the plasmon-assisted harmonic generation, sum and
difference frequency generation, and four-wave mixing of 
graphene nanostructures
\cite{Nat.Comm._5_5725_2014_Cox,Phys.Rev.B_96_045442_2017_Cox}, and these calculated responses can be several order of magnitude larger than that of metal nanoparticles with similar sizes.
Karimi {\it et al.} \cite{Phys.Rev.B_97_245403_2018_Karimi} investigated the Kerr nonlinearity and third harmonic generation of GNR modulated by
scatterings. Attaccalite {\it et al.}\cite{Phys.Rev.B_95_125403_2017_Attaccalite} showed the
importance of excitonic effects in the
third harmonic generation. Wang and Andersen studied the third
harmonic generation of aGNR in the Terahertz frequencies
\cite{J.Phys.DAppl.Phys._49_1_2016_Wang,J.Phys.Condens.Matter_28_475301_2016_Wang,Phys.Rev.B_93_235430_2016_Wang}.  Salazar {\it et al.}
\cite{Phys.Rev.B_93_075442_2016_Salazar} studied two color coherent
control of zGNR, and found that the edge states play an important role
for low photon energies. Recently, Wu {\it et al.} indicated the
importance of the edge states in high-order harmonic generation of
zGNR \cite{Chin.Opt.Lett._18_103201_2020_Wu}. Bonabi and Pedersen
\cite{Phys.Rev.B_99_045413_2019_Bonabi} studied the electric field
induced second harmonic generation of aGNR.

In this paper, we theoretically study the one-color optical injection
current and shift current of zGNR, which are direct currents generated by
light with only one single frequency; they are also widely referred as circularly photogalvanic effects and linear photogalvanic effects. These effects are recently well studied in layered materials including BiFeO$_3$ \cite{BIO} and monolayer Ge and Sn monochalcogenides \cite{PhysRevLett.119.067402}.
Because zGNR possesses the
inversion symmetry, its second order optical responses are forbidden
in the dipole approximation. An external static electric field, which
will be refered as a gate field afterwards, is
applied to break the inversion symmetry. We discuss the dependence of the response coefficients on the gate field strength  and the ribbon width. Our results could be useful for the optoelectronic devices utilizing photogalvanic effects of GNR. 

We arrange the paper as follows. In Section~\ref{sec:model} we
introduce a tight binding model of zGNR with applying a static electric field,
and give the expressions for injection coefficients and shift
conductivities. In Section~\ref{sec:discuss} we discuss the contributions from the edge bands by a simple non-perturbative treatment. In
Section~\ref{sec:width} we discuss the effect of the ribbon width on
these coefficients. We conclude in Section~\ref{sec:conclusion}.

\section{Models}\label{sec:model}
\subsection{Tight-Binding model for electronic states}
\begin{figure*}[!ht]\centering 
\includegraphics[width=7cm]{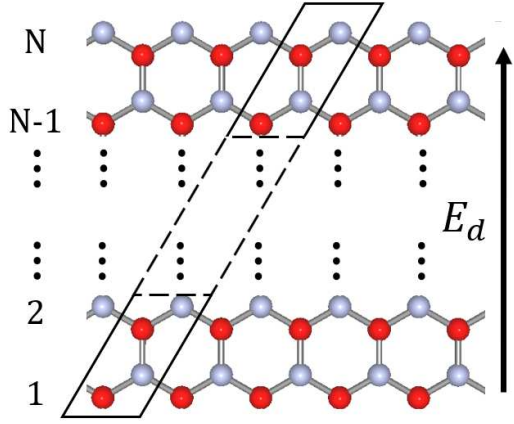}
\caption{Illustration of a N-zGNR. Red
  and gray dots correspond to carbon atoms at the A and B sites,
  respectively. The unit cell of the ribbon is indicated by the
  parallelogram. An external static electric field $E_d$ is applied
  across the ribbon.} 
\label{fig:plot}
\end{figure*}
A zGNR with N zigzag lines (N-zGNR) is illustrated in
Fig.~\ref{fig:plot}. Taking the $x$ axis along the zigzag direction and
the $y$ axis along the perpendicular armchair direction with origin in
the center of the ribbon, the carbon
atoms locate at ${\bm R}_{nm\alpha} = n{\bm a} + (m-1){\bm a}_2 + {\bm
  \tau  }_\alpha-\hat{\bm y}W/2$, 
where ${\bm a}=a_0\hat{\bm x}$ is the primitive lattice
vector with the lattice constant ${a}_0  = 2.46$ \AA, 
${\bm a}_2=a_0(\hat{\bm x}+\sqrt{3}\hat{\bm y})/2$ and $m=1,2,...,N$
labelling zigzag lines, and 
${\bm \tau}_\alpha$ with $\alpha=A,B$ gives different atom sites as ${\bm \tau}_A=0$ and
${\bm \tau}_B=({\bm  a}_1+{\bm a}_2)/3$. The width of a N-zGNR is
$W=(N-2/3)\sqrt{3}a_0/2$ by taking as the distance between the outermost A and B atom lines. We describe the electronic states in a tight-binding
model formed by carbon $2p_z$ orbitals with considering the nearest
neighbor coupling only. When a gate field $E_d$ is applied, the unperturbed Hamiltonian can be written as 
\begin{equation}
  \hat{H}_0=\hat{H}_h-eE_d\hat{y}
\end{equation}
with the electron charge $e=-|e|$. The first term $\hat {H}_h$ is a hopping term with matrix elements
\begin{align}
  _t\left\langle
    n_{1}m_{1}A\right|\hat{H}_h\left|n_{2}m_{2}B\right\rangle_t&={_t\left\langle
    n_{1}m_{1}A\right |\hat{H}_h\left|n_{2}m_{2}B\right\rangle_t}^*\notag\\
    &=-\gamma_0(\delta_{n_1,n_2}\delta_{m_1,m_2}+\delta_{n_1,n_2}\delta_{m_1,m_2+1}+\delta_{n_1+1,n_2}\delta_{m_1,m_2})\,,\\
  _t\left\langle n_{1}m_{1}\alpha\right|\hat{H}_h\left |n_{2}m_{2}\alpha\right\rangle_t&=0\,,
\end{align}
where $\gamma_0=2.7\,$eV is a hopping parameter between nearest
neighbours, the ket $|nm\alpha\rangle_t$ stands for the electronic state
of the $2p_z$ orbital of the carbon atom located at ${\bm
  R}_{nm\alpha}$. The second term is the electrostatic potential, $\hat{y}$ is the
$y$-component of the position operator $\hat{\bm r}$. In this model, the in-plane
position operator has nonzero matrix elements only at the same site as
\begin{equation}
    _t\left\langle n_{1}m_{1}\alpha_{1}\right|\hat{\bm
      r}\left|n_{2}m_{2}\alpha_{2}\right\rangle_t={\bm R}_{n_{1}m_{1}\alpha_{1}}\delta_{n_1n_2}\delta_{m_1m_2}\delta_{\alpha_1\alpha_2}\,.
\end{equation}
In Bloch states basis formed by
\begin{equation}
    \left|m\alpha,k\right\rangle_b =\sqrt{\frac{a_0}{2\pi}} \sum_n
    e^{ina_0k}\left|nm\alpha\right\rangle_t\,, \text{for } 0\leq k < g\,,
  \end{equation}
  with $g={2\pi}/{a_0}$ being the width of the Brillouin zone,  the matrix elements of the Hamiltonian $\hat{H}_0$,
  position operator $\hat{\bm r}$,
  and velocity operator $\hat{\bm v}=[\hat{\bm r},\hat{H}_0]/(i\hbar)$ become
\begin{align}
  _b\langle
  m_1\alpha_1,k_1|\hat{H}_0|m_2\alpha_2,k_2\rangle_b&=\tilde H^0_{m_1\alpha_1,m_2\alpha_2;k}\delta(k_1-k_2)\,,\\
_b\langle
  m_1\alpha_1,k_1|\hat{\bm r}|m_2\alpha_2,k_2\rangle_b&=\left[\tilde{\bm
                                                        r}_{m_1\alpha_1,m_2\alpha_2;k_1}+i\hat{\bm x}\frac{\partial}{\partial k_1}\right]\delta(k_1-k_2)\,,\\
  _b\langle m_1\alpha_1,k_1|\hat{\bm
  v}|m_2\alpha_2,k_2\rangle_b&=\tilde{\bm v}_{m_1\alpha_1,m_2\alpha_2;k}\delta(k_1-k_2)\,.
\end{align}
The quantities $\tilde P_{m_1\alpha_1,m_2\alpha_2;k}=\sum_n e^{ina_0k}
  {_t\langle}nm_1\alpha_1|\hat{P} | 0m_2\alpha_2\rangle_t$ for $P= H^0$, $\bm
r$ and $\bm v$ are the Fourier transform of their matrix elements in
the tight binding orbitals, and their matrix elements are 
\begin{align}
  \tilde H^0_{m_1A,m_2B;k}&\equiv [\tilde H^0_{m_2B,m_1A;k}]^\ast
                          =\gamma_0(1+e^{ika_0})\delta_{m_1m_2} +
                                 \gamma_0\delta_{m_1+1,m_2}\,,\\
  \tilde H^0_{m_1\alpha,m_2\alpha;k}&=-eE_d\tilde r^y_{m_1\alpha,m_2\alpha;k}\,,
\end{align}
and
\begin{align}
  \tilde{\bm r}_{m_1\alpha_1,m_2\alpha_2;k}&=\delta_{m_1m_2}\delta_{\alpha_1\alpha_2}\bm  R_{0m_1\alpha_1}\,,\\
  \tilde{\bm v}_{m_1\alpha_1,m_2\alpha_2;k}
  &= \frac{1}{i\hbar}[\tilde{\bm r}_k,
    \tilde H^0_k]_{m_1\alpha_1,m_2\alpha_2}
    +\hat{\bm x} \frac{1}{\hbar}\frac{\partial}{\partial k} \tilde H^0_{m_1\alpha_1,m_2\alpha_2;k}\,.
\end{align}
In the last equation $\tilde{\bm r}_k$ and $\tilde H^0_k$ are treated
as matrices with indexes $m\alpha$. The gate field modifies the on-site energy of each atom. 

The band eigenstates $|sk\rangle$ with band index $s$ can be written as 
\begin{align}
  |s k\rangle = \sum_{m\alpha} [C_{sk}]_{m\alpha}|m\alpha,k\rangle_b\,.
\end{align}
where the coefficients $C_{sk}$ are column eigenvectors satisfying
\begin{equation}
  \tilde H^0_kC_{sk}=\varepsilon_{sk}C_{sk}\,,
\end{equation}
with the corresponding eigen energy $\varepsilon_{sk}$.

For optical response, the most important quantity is the Berry
connection $\bm\xi_{s_1s_2k}$ between band eigenstates, which is
defined as 
  \begin{align}
    \bm \xi_{s_1s_2k}&=C^{\dag}_{s_1k} \left(\tilde{\bm r}_k + i\hat{\bm x}
  \frac{\partial}{\partial k}\right)C_{s_2k}\,.
\end{align}
The term $\xi^y_{s_1s_2k}$ can be evaluated directly. However, due to
the derivative with respect to $k$, the values of $\xi^x_{s_1s_2k}$
depend on the phase of  the eigen vectors $C_{sk}$ and is not easy to
be evaluated directly. Usually the off-digonal terms can be 
evaluated from the matrix elements of velocity operator
\begin{align}
  \bm v_{s_1s_2k} = C_{s_1k}^\dag\tilde{ \bm v}_k C_{s_2k}\,.
\end{align}
The usually used quantities are $\bm r_{\bm k}$, which are defined as 
\begin{align}
  r_{s_1s_2k}^y&= \xi_{s_1s_2k}^y\,,\text{ for all } s_1,s_2\,,\\
  r_{s_1s_2k}^x &\equiv \begin{cases} \xi^x_{s_1s_2k} =
    \frac{v^x_{s_1s_2k}}{i\omega_{s_1s_2k}} & \text{ for } s_1\neq
    s_2\,,\\
    0& \text{ for } s_1=s_2\end{cases}
\end{align}
with
$\hbar\omega_{s_1s_2k}=\varepsilon_{s_1k}-\varepsilon_{s_2k}$. The
digonal term of $\xi_{ssk}^x$ appears in terms
\begin{align}
  {\cal R}_{s_1s_2k}^{cx} = \frac{\partial }{\partial k}r_{s_1s_2k}^c -
  i(\xi_{s_1s_1k}^x-\xi_{s_2s_2k}^x)r_{s_1s_2k}^c\,, \text{ for }
  s_1\neq s_2\,.
\end{align}
with the Roman letter $c$ in the superscript standing for the Cartesian directions $x$ or $y$. A direct calculation
gives
\begin{align}
  {\cal R}_{s_1s_2k}^{cx}
    &=-\frac{\Delta_{s_1s_2k}^xr_{s_1s_2k}^c}{\omega_{s_1s_2k}}
      +\frac{i[r_k^x,v_k^c]_{s_1s_2}+M_{s_1s_2k}^{cx}}{i\omega_{s_1s_2k}}\,,
\end{align}
with $\Delta_{s_1s_2k}^b=v_{s_1s_1k}^b-v_{s_2s_2k}^b$  and 
\begin{align}
  M_{s_1s_2k}^{cx} &= C_{s_1k}^\dag \left(\frac{\partial}{\partial
                     k}\tilde v_k^c-i[\tilde
                     r^x_k,\tilde
                     v_k^c]\right)C_{s_2k}\,.
\end{align}
For zGNR, ${\cal R}^{yx}_{s_1s_2k} = i[r_k^x,r_k^y]_{s_1s_2}$.

For very narrow zGNR with $N<30$, the interaction between carriers at both edges plays an
important role to form antiferromagnetic order, for which the spin
orientations are opposite for different edges. For wide ribbons
$N>30$, the ferromagnetic-antiferromagnetic energy differences per
unit cell are reduced below the order of 1
meV\cite{PhysRevLett.97.216803}, hence the magnetic order can be ignored.

\subsection{Injection currents and shift currents}
In this work, we are interested in the shift current and one-color
injection current, both of
which arise from the second order optical response. For an incident electric
field $\bm E(t)=\bm E_0(t) e^{-i\omega t} + c.c.$ with the slow
varying envelope function $\bm E_0(t)$, the response current
includes a (quasi) dc current component $\bm J_0(t)=J_0(t)\hat{\bm x}$, which is along
the ribbon extension direction only because a dc current cannot flow along the confined
dimension. This current approximately includes two parts $J_0(t) =
J_i(t) +  J_s(t)$. 
The first term $J_i(t)$ is a one-color injection current, and it is
\begin{align}
  \frac{d}{dt}J_i(t) = 2i\eta^{xbc}(\omega) E_0^b(t) \left[E_0^c(t)\right]^\ast\,,
\end{align}
and the effective sheet injection rate is $\eta^{xbc}(\omega)
= \sum_{s_1s_2}\eta^{xbc}_{s_1s_2}(\omega)$ with
\begin{align}
  \eta_{s_1s_2}^{xbc}(\omega)&=-\frac{i\pi e^3}{W\hbar^2}\int
  \frac{dk}{2\pi}\Delta^x_{s_{1}s_{2}k}\left(
  r^{c}_{s_{2}s_{1}k}r^{b}_{s_{1}s_{2}k}-r^b_{s_2s_1k}r^c_{s_1s_2k}\right)f_{s_{2}s_{1}k}\delta\left(\omega_{s_{1}s_{2}k}-\omega\right)\,.\label{eq:eta}
\end{align}
Here $f_{s_2s_1k}=f_{s_2k}-f_{s_1k}$ gives the population difference in
 two states $|s_2k\rangle$ and $|s_1k\rangle$, and $f_{sk} =
 [1-e^{(\varepsilon_{sk}-\mu)/k_BT}]^{-1}$ is Fermi-Dirac
 distribution for chemical potential $\mu$ and temperature $T$.  The spin degeneracy
 has been included in Eq.~(\ref{eq:eta}). 
The second term $J_s(t)$ is a shift current, and it is
\begin{align}
  J_s(t) &= 2\sigma^{xbc}(\omega) E_0^b(t) \left[E_0^c(t)\right]^\ast\,,
\end{align}
where the effective sheet shift conductivity is $\sigma^{xbc}(\omega)
= \sum_{s_1s_2}\sigma_{s_1s_2}^{xbc}(\omega)$ with
\begin{align}
  \sigma_{s_1s_2}^{xbc}(\omega)&=-\frac{i\pi
  e^3 }{W\hbar^2}\int\frac{dk}{2\pi}f_{s_{2}s_{1}k}\left(r^{b}_{s_{1}s_{2}k}{\cal
  R}^{cx}_{s_{2}s_{1}k}+r^{c}_{s_{1}s_{2}k}{\cal R}^{bx}_{s_{2}s_{1}k}\right)\delta(\omega_{s_{1}s_{2}k}-\omega)\,.\label{eq:sigmas}
\end{align}

Here we briefly discuss the general properties of $\eta^{xbc}(\omega)$ and $\sigma^{xbc}(\omega)$ from the
symmetry argument. The response coefficients of
$\eta^{xbc}(\omega)$ and $\sigma^{xbc}(\omega)$ are third order
tensors. As a static electric field is applied along the
$y$-direction, a zGNR possesses a symmetry $x\leftrightarrow
-x$ and the time reversal symmetry. We list the results for
$A^{xbc}$ ($A\to \eta$ or $\sigma_s$) under each symmetry operation: (1) The symmetry $x\leftrightarrow
-x$ determines that the nonzero components are $A^{xxy}$ and
$A^{xyx}$. (2) A direct observation of Eqs.~(\ref{eq:eta}) and
(\ref{eq:sigmas}) gives $\eta^{xbc}(\omega)=-\eta^{xcb}(\omega)$ and
$\sigma_s^{xbc}(\omega)=\sigma_s^{xcb}(\omega)$. (3) The time reversal symmetry\cite{PRB_61_5337_Sipe} gives $\bm r_{s_1s_2k} =
\bm r_{s_2s_1(-k)}=[\bm r_{s_1s_2(-k)}]^\ast$,  $\bm v_{s_1s_2k}=-\bm
v_{s_2s_1(-k)}=-[\bm v_{s_1s_2(-k)}]^\ast$, and
$\varepsilon_{sk}=\varepsilon_{s(-k)}$. Furthermore, we can derive
$\Delta^a_{s_1s_2k}=-[\Delta_{s_1s_2(-k)}^a]^\ast$ and ${\cal
  R}^{cx}_{s_1s_2k}=-{\cal R}^{cx}_{s_2s_1(-k)}=-[{\cal R}^{cx}_{s_1s_2(-k)}]^\ast$. Then we get
$\eta^{xbc}(\omega)=[\eta^{xbc}(\omega)]^\ast$ from
Eq.~(\ref{eq:eta}) and $\sigma^{xbc}(\omega)=[\sigma^{xbc}(\omega)]^\ast$
from Eq.~(\ref{eq:sigmas}). Using the operations (1)-(3) we find the
nonzero components $\eta^{xxy}(\omega)=-\eta^{xyx}(\omega)$ and $\sigma_s^{xxy}(\omega)=\sigma_s^{xyx}(\omega)$ are real
numbers. 

Explicitly, by taking the light fields as $\bm
E_0(t)=E_0(t)\begin{pmatrix}\cos\theta\\ e^{i\phi}\sin\theta
\end{pmatrix}$, the injection and shift currents can be written as
\begin{align}
  \frac{d}{dt}J_i(t) 
    &= 4\eta^{xxy}(\omega)[E_0(t)]^2\cos\theta\sin\theta\sin\phi\,,\\
  J_s(t) 
    &= 4\sigma^{xxy}(\omega)[E_0(t)]^2\cos\theta\sin\theta\cos\phi\,.
\end{align}
Here $\theta$ and $\phi$ are the polarization orientation angles with respect to the direction
$\hat{\bm x}$ and the circularity, respectively. Therefore, the appearance of these currents requires
  both the $x$ and $y$ components of the electric field. The
  circularly polarized light ($\phi=\pi/2$) generates injection currents only,
  while the linearly polarized light ($\phi=0$) generates shift currents only.

\section{Result and Discussions}\label{sec:discuss}

\subsection{Band structure}
\begin{figure*}[!ht]\centering
\includegraphics[height=7.5cm]{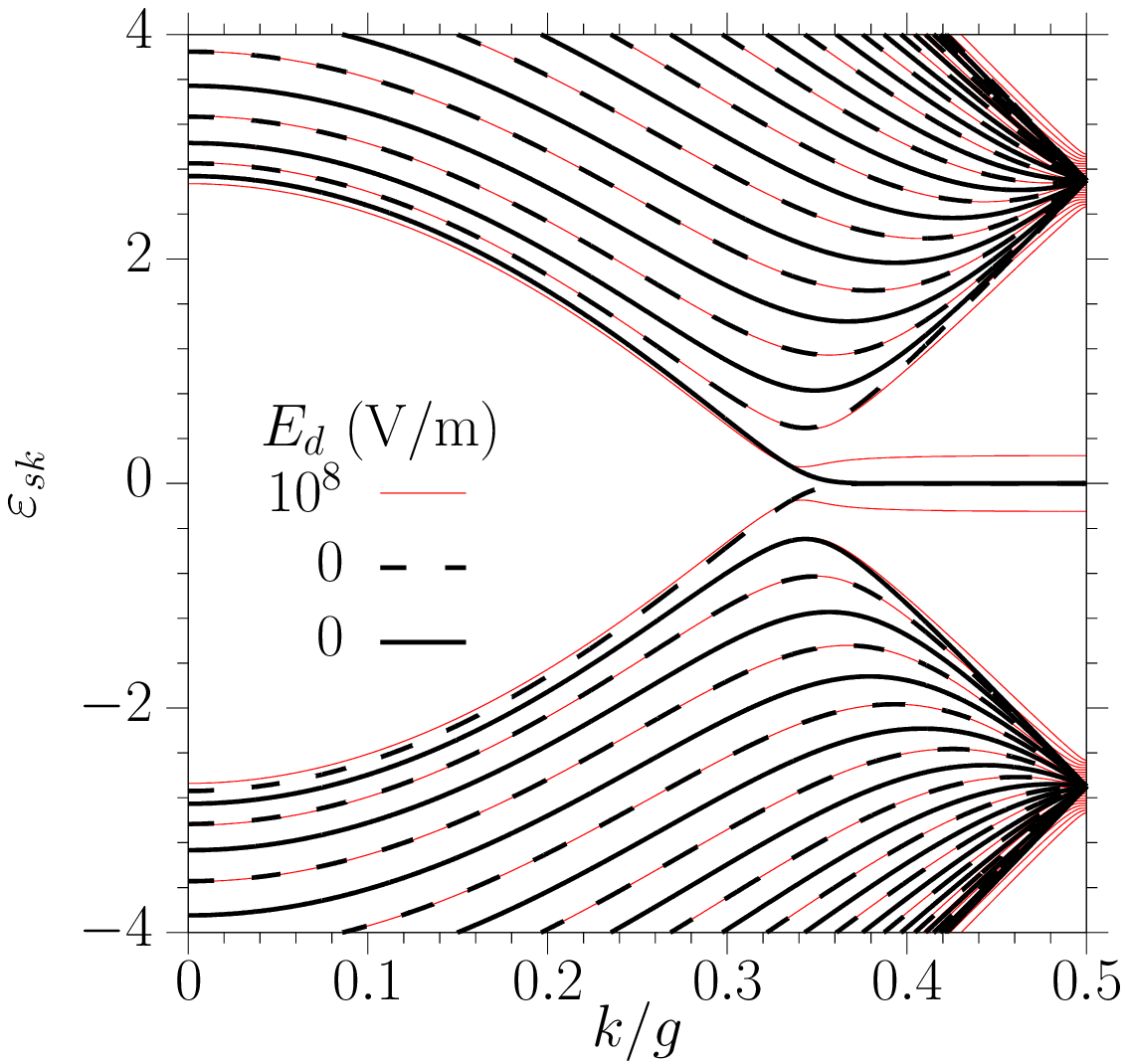}
\includegraphics[height=7.5cm]{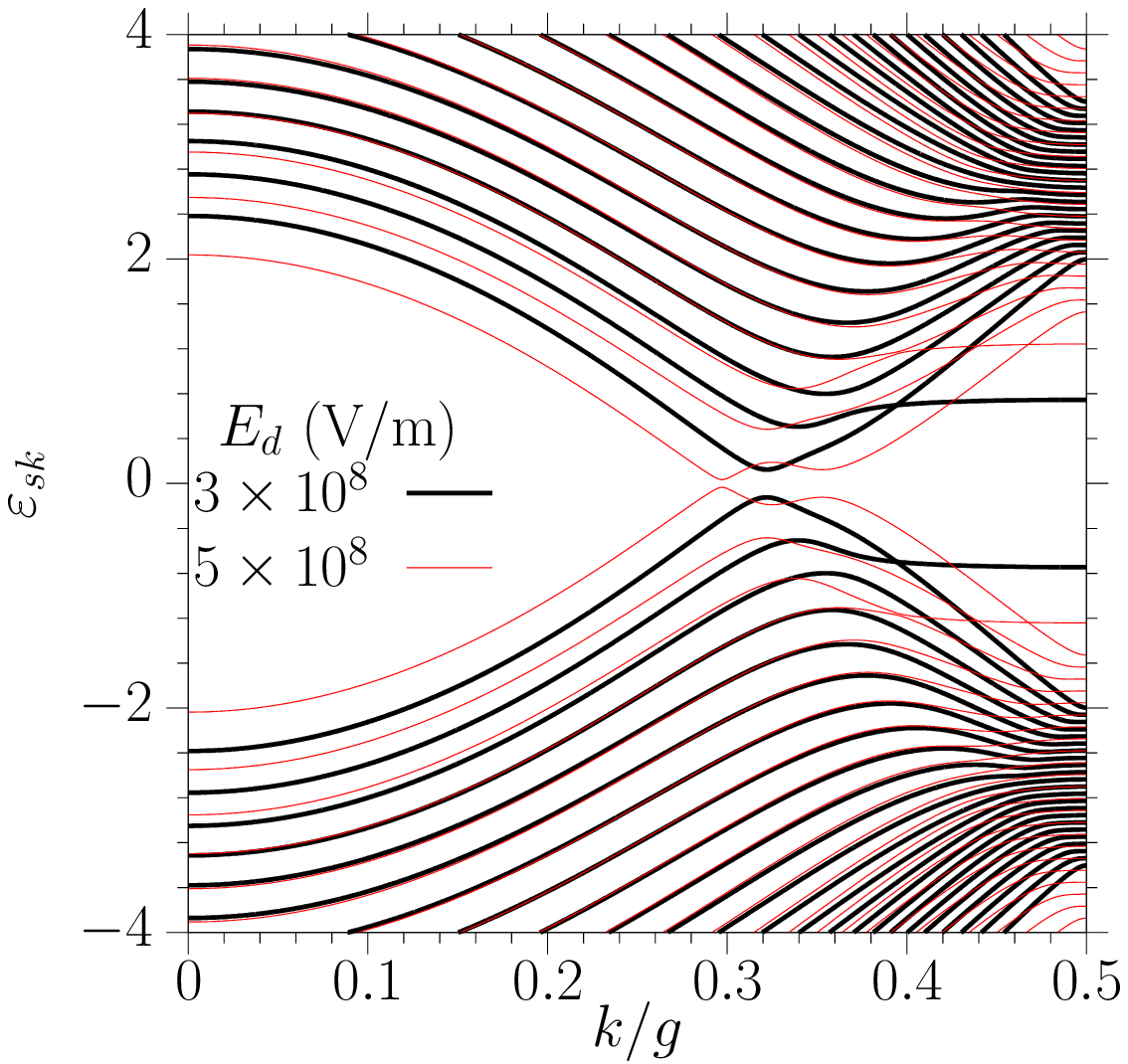}\\
\includegraphics[height=7.5cm]{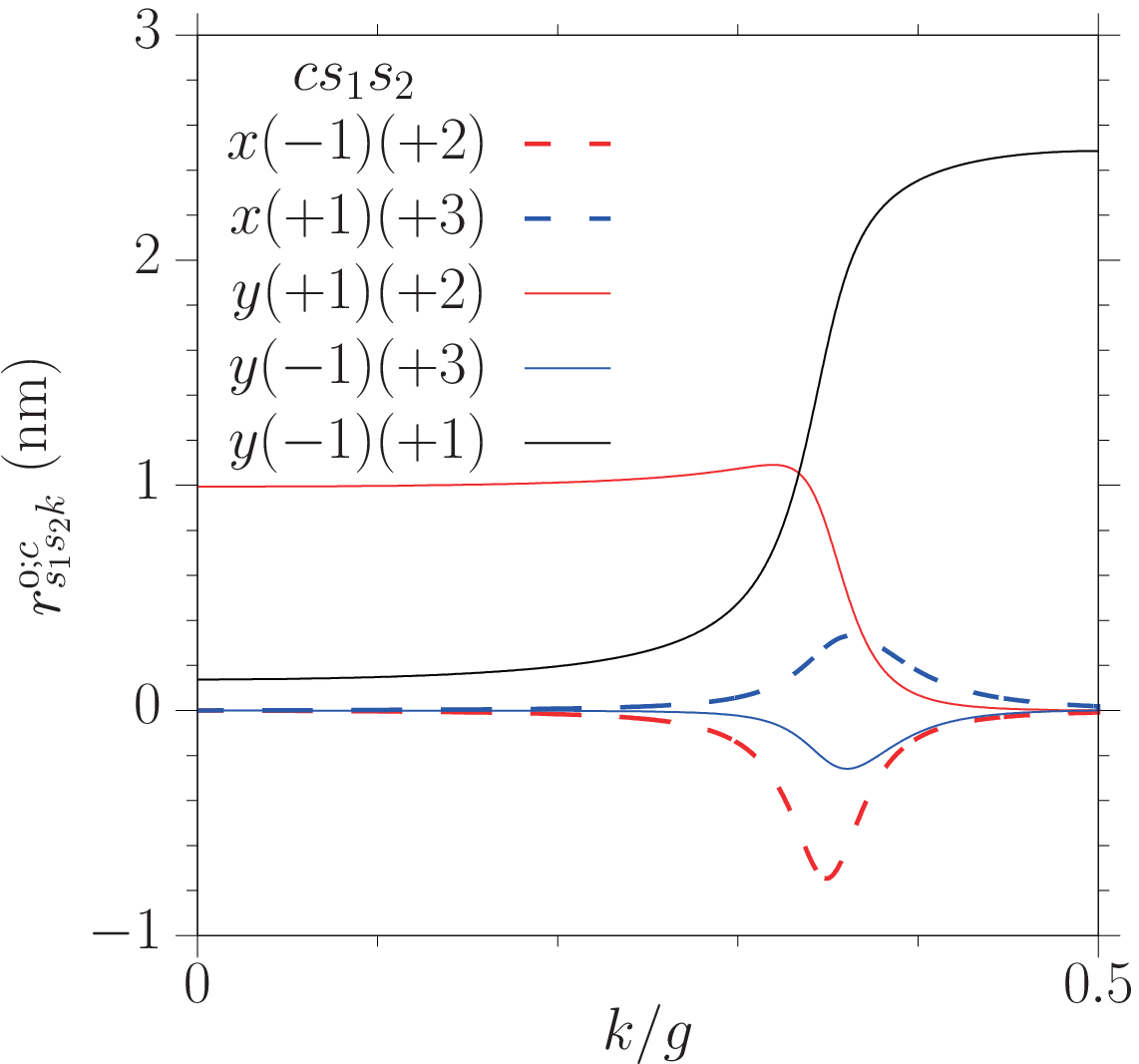}
\includegraphics[height=7.5cm]{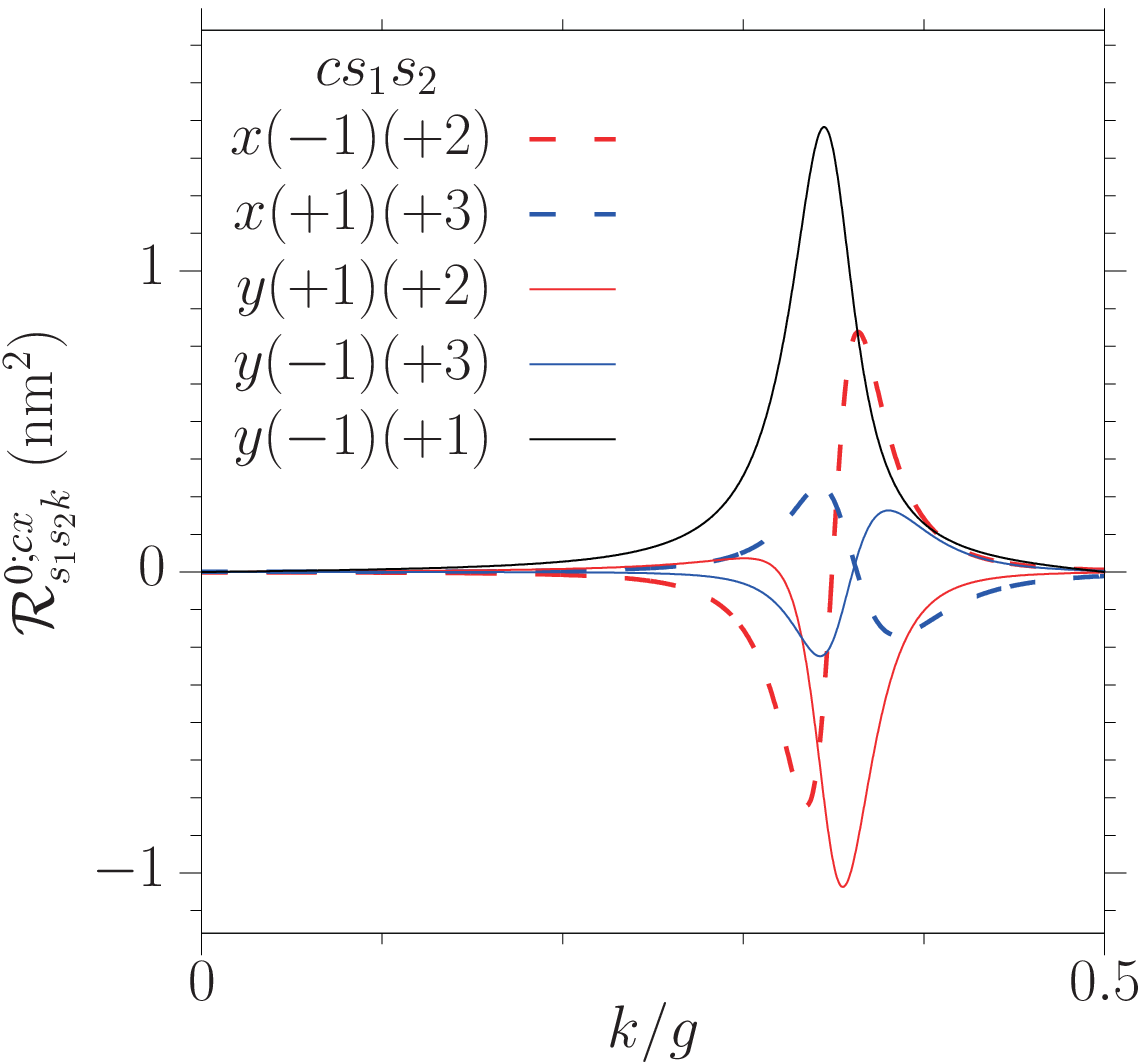}
\caption{(a,b) Band structures of 24-zGNR for gate field $E_d=0$,
  $10^8$~V/m, $3\times 10^8$~V/m, and $5\times 10^8$~V/m. At zero
  field, the dashed and solid curves correspond to different
  parity. The  matrix elements of (c) $r^{0;c}_{s_1s_2k}$ and  (d) ${\cal
    R}^{0;cx}_{s_1s_2k}$ at zero gate field, with solid (dashed) curves
  for imaginary (real) parts.
}\label{fig:bands}
\end{figure*}
We illustrate the band structures of a 24-zGNR ($W\approx5$~nm) for
different gate field $E_d$ in Fig.~\ref{fig:bands} (a,b). The bands 
with energies higher than zero are labelled by $s=+1,+2,\cdots$
successively from low energy band to high energy band, and
those with energy lower than zero are labelled by $s=-1,-2,\cdots$ in
a mirror way. From the symmetry $x\to-x$, the band energies satisfy
$\varepsilon_{sk}=\varepsilon_{s (g-k)}$ and $\varepsilon_{s
  k}=-\varepsilon_{(-s) k}$, and thus they are shown only in half of
Brillouin zone.   
The band structure at zero gate field is plotted in
Fig.~\ref{fig:bands} (a) as black solid and dashed curves. Two bands
$s=\pm 1$ are almost flat in the middle of the Brilluion zone,
indicating the edge states. The energy difference
$\varepsilon_{(+1)k}-\varepsilon_{(-1)k}$ decreases as $k$
approaching $g/2$ and becomes less than
1~meV for $0.38g<k<0.62g$. At $k=g/2$, the two states are strictly
degenerate. All other electronic states are confined states. At
$k=g/2$, all the state $|sg/2\rangle$ for $s>1$ are degenerate at
energy $\gamma_0$, and all states $|sg/2\rangle$ for $s<-1$ are
degenerate at energy $-\gamma_0$. At zero gate field, the
inversion symmetry is preserved, and the parity is a good quantum
number for each band as $\zeta_s=(-1)^{s+1}\text{sgn}[s]$
\cite{PhysRevB.95.155438,Phys.Rev.B_93_075442_2016_Salazar}, which is shown in dashed and solid
curves in Fig.~\ref{fig:bands} (a). There exist selection rules for
the velocity matrix elements as $v^x_{s_1s_2k} = 0$ for
$\zeta_{s_1}\neq \zeta_{s_2}$ and $v^y_{s_1s_2k} = 0$ for $\zeta_{s_1}
= \zeta_{s_2}$, and the same selection rules hold for $\bm \xi_{s_1s_2k}$. 
Therefore, the nonzero $\xi^y_{s_1s_2k}$ between bands with different
parities indicates that the gate field can couple bands with different parities and then the band parity is no longer a good quantum number.

Figure~\ref{fig:bands} (c) gives the $k$-dependence of $r^{0;c}_{s_1s_2k}$
for different sets of $cs_1s_2$, where a quantity at zero gate field
is indicated by a superscript ``0''. With choosing the wave functions
appropriately, $r^{0;x}_{s_1s_2k}$ can be set as pure imaginary numbers and
$r^{0;y}_{s_1s_2k}$ as real numbers. For $r^{0;y}_{(-1)(+1)k}$, it is
close to a value $W/2=2.5$~nm for edge states, and decreases  for
confined states ($k<0.34g$) as $k$ decreases to
0. Figure~\ref{fig:bands} (d)  gives the $k$-dependence of ${\cal
  R}^{0;cx}_{s_1s_2k}$ for the same sets of $cs_1s_2$, which locates
at around $k\sim0.33g$. We have also compared the values $\partial_k
r^{0;c}_{s_1s_2k}$ and ${\cal R}^{0;cx}_{s_1s_2k}$, and they show
negligible difference which indicates all $\xi_{ssk}^{0;x}$ can be
taken as zero, as used in Appendix A. 

The band structure at a gate field $E_d= 10^8$~V/m is also plotted in
Fig.~\ref{fig:bands}(a). Such gate field mostly affects the bands
$s=\pm1$. It opens the degenerate point at $k=g/2$ to an energy difference $|e|E_0W\sim 0.5$~eV,
and separates the two nearly degenerate flat bands with
energies around $\pm0.25$~eV. The gap of these two bands is about   $\sim0.3$~eV located at $k\sim0.34g$. The band structures at stronger gate field $E_d=3\times 10^8$~V/m and
$5\times 10^8$~V/m are shown in Fig.~\ref{fig:bands} (b). In both
cases, the gate fields can significantly affect more bands including
$s=\pm2$ and $s=\pm3$. When the field strength $E_d$ is large enough,
the gap can be closed again, and all bands are significantly modified. In this work, we limit the gate field $E_d<10^8$~V/m to ensure the reasonableness of our tight binding model.

To better understand the effects of a weak gate field on the edge states,
we present a simple two band model. The sub-Hilbert space is formed by
$\{|(+1)k\rangle^0,  |(-1)k\rangle^0\}$. The Hamiltonian in this subspace is 
\begin{equation}
  H^{edge}_k =   \begin{pmatrix} 
    \epsilon_k & d_k \\ d_k & -\epsilon_k
  \end{pmatrix} \label{eq:smodel}
\end{equation}
where  $\epsilon_k=\varepsilon_{(+1)k}^0$ is the energy of
band "+1" at zero gate field, and $d_k= |e|E_dr^{0;y}_{(+1)(-1)k}$ is the
coupling strength which can be chosen as a real positive number. We
have used $\xi_{ssk}^{0;y}=0$ to obtain Eq.~(\ref{eq:smodel}). From
Fig.~\ref{fig:bands}(c) the matrix element of $r^{0;y}_{(+1)(-1)k}$ is
around $W/2$ for the edge states $k\sim g/2$, but decreases as k moves to 0. 
The Hamiltonian in Eq.~(\ref{eq:smodel}) has the eigenstates 
\begin{align}
  |sk\rangle&=\frac{1}{\sqrt{2}}\left[s\sqrt{1+sN_k} |(+1)k\rangle^0
              +\sqrt{1-sN_k} |(-1)k\rangle^0\right]\,, &\text{ for } s =\pm1,
\end{align}
and the eigenenergies
\begin{align}
  \varepsilon_{sk}&=s\sqrt{\epsilon_k^2+d_k^2}\,,
\end{align}
with $N_k=\epsilon_k/\varepsilon_{(+1)k}$. For edge states
at $k=g/2$, $\epsilon_k=0$ and $\varepsilon_{sk}=s|e|E_dW/2$; as $k$
moving towards $0$, $\epsilon_k$ increases slowly till $k<g/3$
but $d_k$ decreases quickly, which gives a dip in the spectra of
$\varepsilon_{+k}$ around $k\sim g/3$; when $k$ further moving, the
bands $s=\pm1$ are no longer nearly degenerate, and the effect of the
gate field can be treated as a perturbation.

The effects of the gate field on higher bands are basically perturbative, thus to focus on the influence of the edge states,  we calculate the Berry connections of the electronic states $\{ |sk\rangle, |lk\rangle^0; s=\pm 1, l\neq \pm 1\}$  as
\begin{align}
  \bm \xi_{slk} &= \frac{1}{\sqrt{2}}
                  \left[s\sqrt{1+sN_k}\bm\xi_{(+1)lk}^0+\sqrt{1-sN_k}\bm\xi_{(-1)lk}^0\right]\,,\\
 \bm
  \xi_{(+1)(-1)k}&=\frac{1}{2}\frac{i\partial_kN_k}{\sqrt{1-N_k^2}}\hat{\bm
                   x} + N_k\xi_{(+1)(-1)k}^{0;y}\hat{\bm y}\,.
\end{align}
A detailed derivation is given in Appendix~\ref{app:berry}.

\subsection{Injection coefficients of $24$-zGNR}
\begin{figure*}[htp]
  \centering
  \includegraphics[height=6cm]{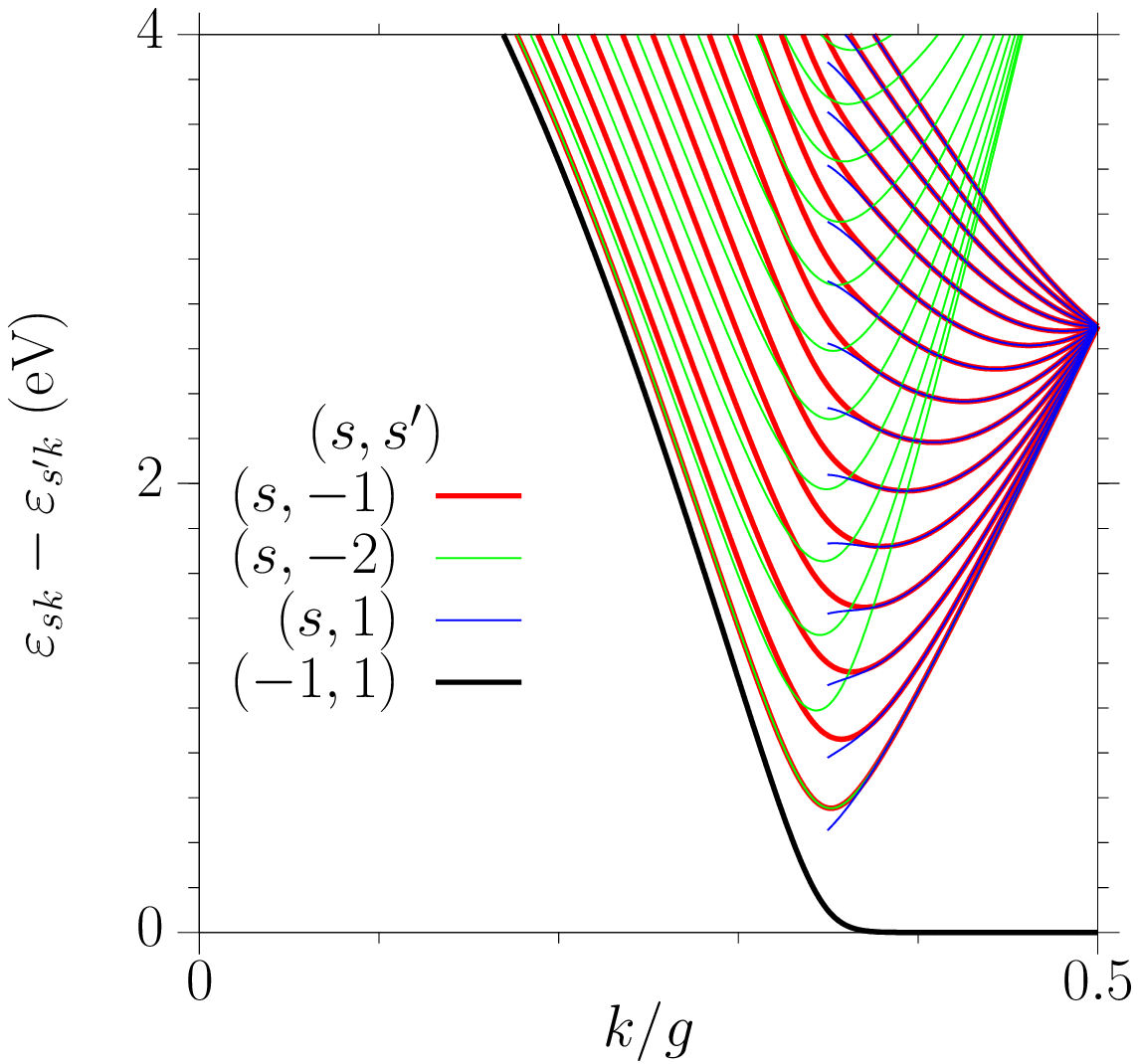}
  \includegraphics[height=6cm,width=6.5cm]{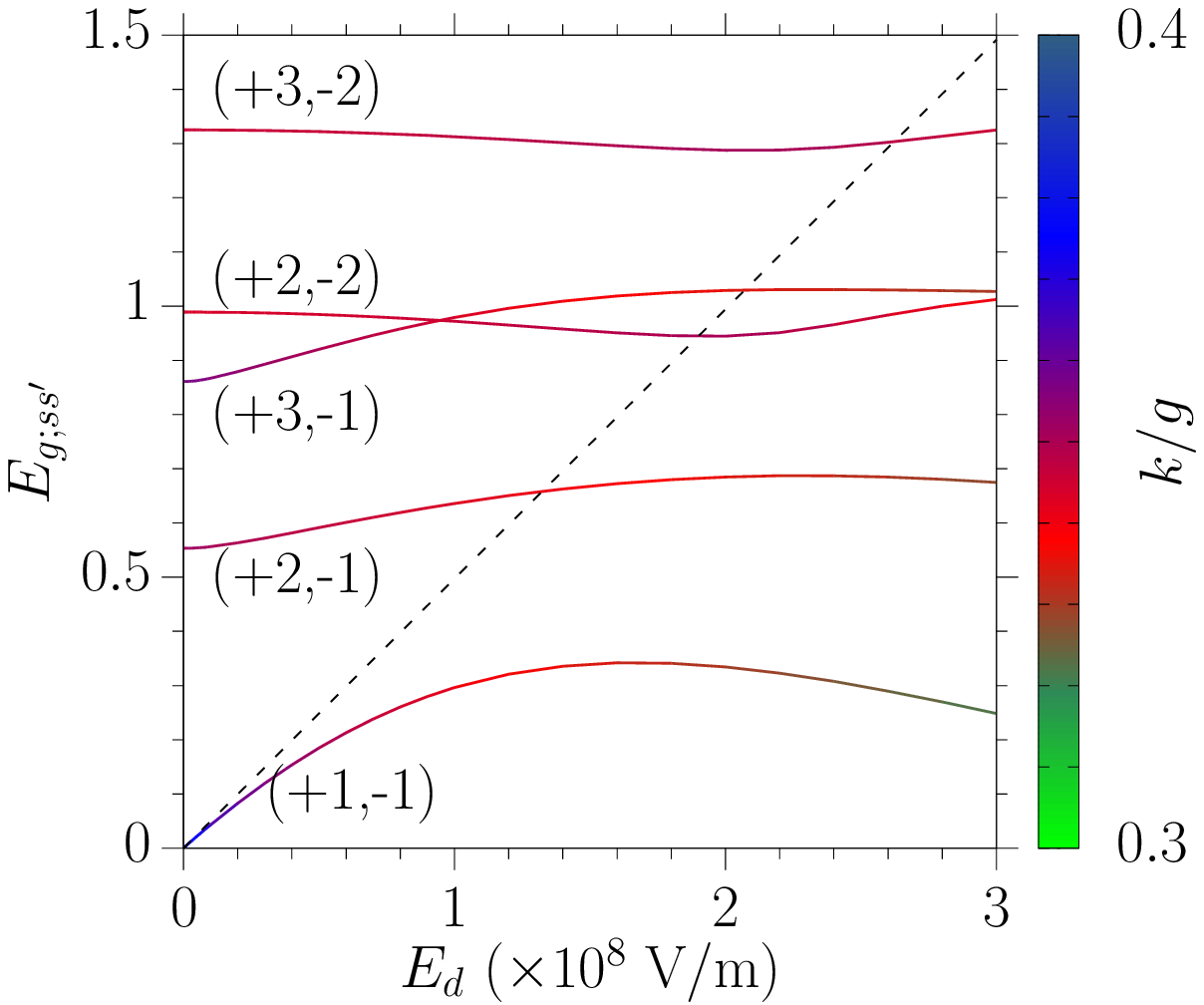}
  \caption{(a) The energy difference
    $\varepsilon_{sk}-\varepsilon_{s^\prime}$ for different 
  $(s,s^\prime)$ pairs at zero gate field. (b) The gate field
  dependence of the energy gaps $E_{g;ss^\prime}$ between different
  bands $(s,s^\prime)$. The line color indicates the $k$ values of these
  gaps. The black dotted line gives the energy difference
  $\varepsilon_{(1)g/2}-\varepsilon_{(-1)g/2}$.}
\label{fig:jdos}
\end{figure*}
We turn to the numerical evaluation of the injection coefficients in Eq.~(\ref{eq:eta}) and the shift
conductivity in Eq.~(\ref{eq:sigmas}). During the numerical evaluation, the Brillouin zone is divided into a $3100$ grid, the $\delta$ function
is approximated by a Gaussian function
\begin{equation}
  \delta(\hbar\omega_{s_1s_2k}-\hbar\omega)  \rightarrow \frac{1}{\sqrt{\pi}\Delta}e^{-(\hbar\omega_{s_1s_2k}-\hbar\omega)^2/\Delta^2}
\end{equation}
with a broadening width $\Delta=2$~meV, and the temperature is chosen at
room temperature. The functions $\delta(\hbar\omega_{ss^\prime k}-\hbar\omega)$
are associated with the joint
density of states, which gives the weight to the
optical transition from the $s^\prime$ band to the $s$ band.
It can be evaluated exactly as
\begin{align}
  \delta(\hbar\omega_{ss^\prime k}-\hbar\omega)   = \sum_j \frac{1}{\hbar|\Delta_{ss^\prime k}|}\delta(k-k_j)\,,\label{eq:deltaexact}
\end{align}
with $k_{j}$ satisfying $\omega_{ss^\prime k_j}=\omega$.
\begin{figure*}[htp]
  \centering
  {\includegraphics[width=7.5cm]{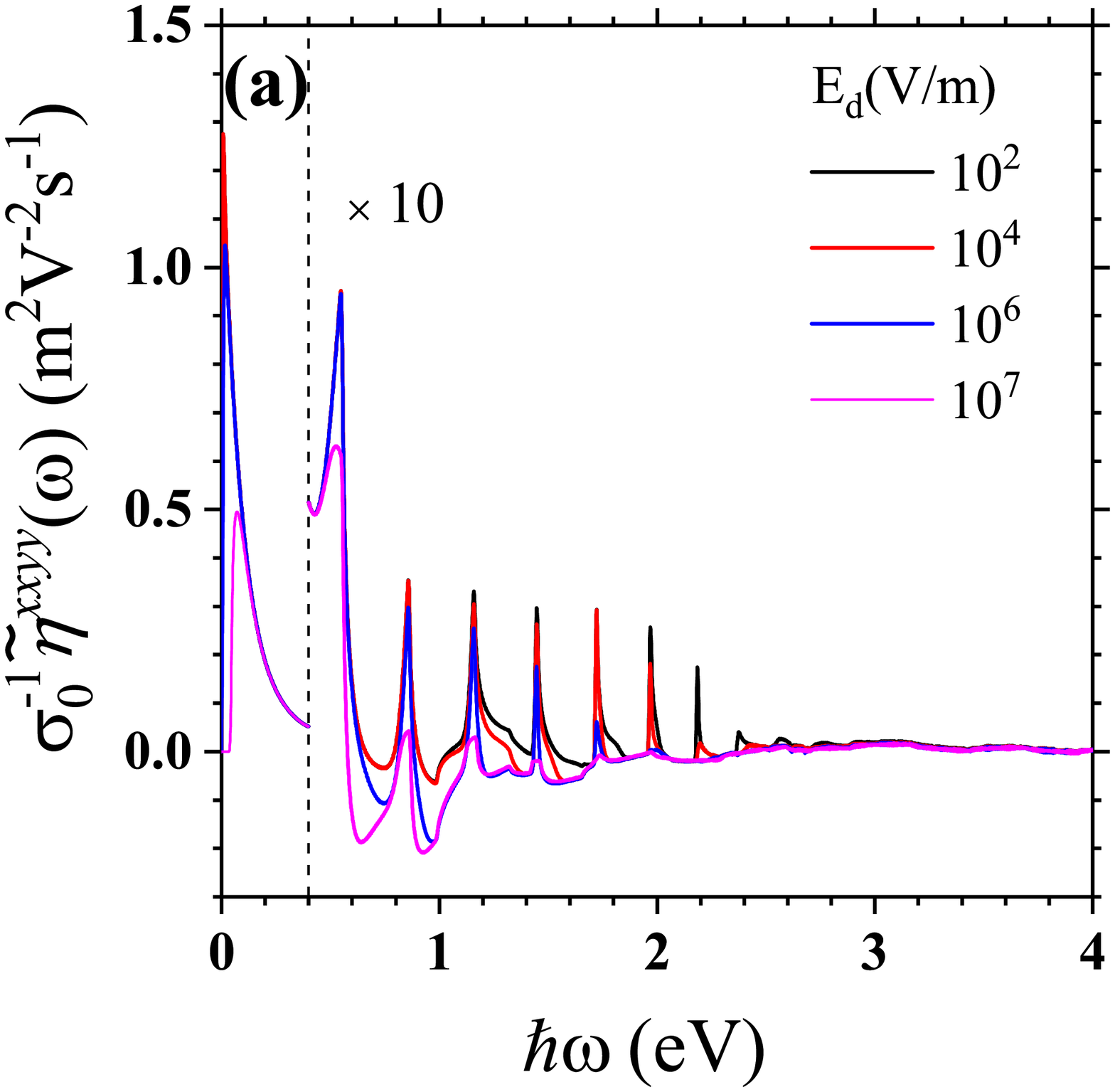}\includegraphics[width=7.5cm]{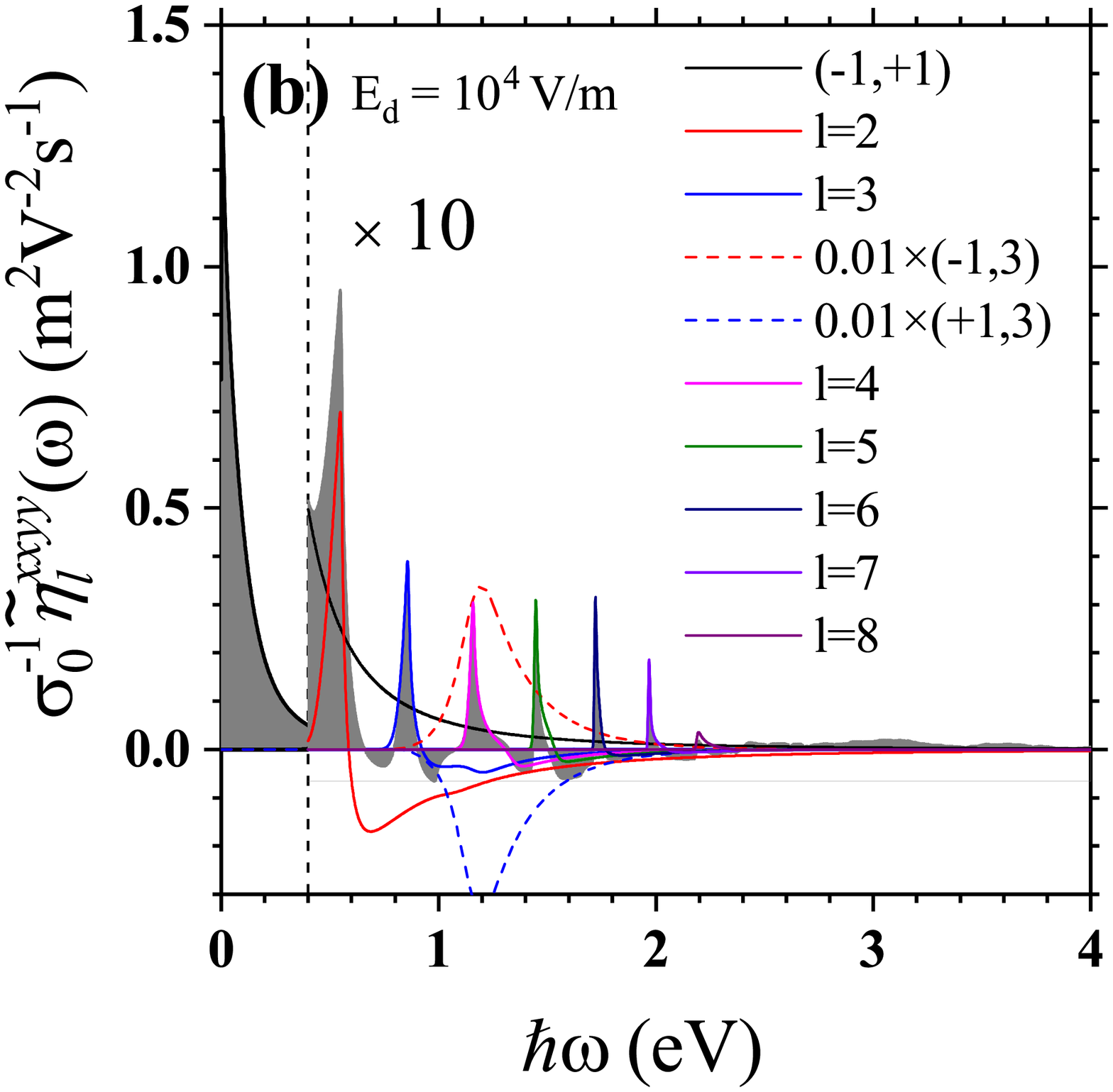}\\
    \includegraphics[width=7.5cm]{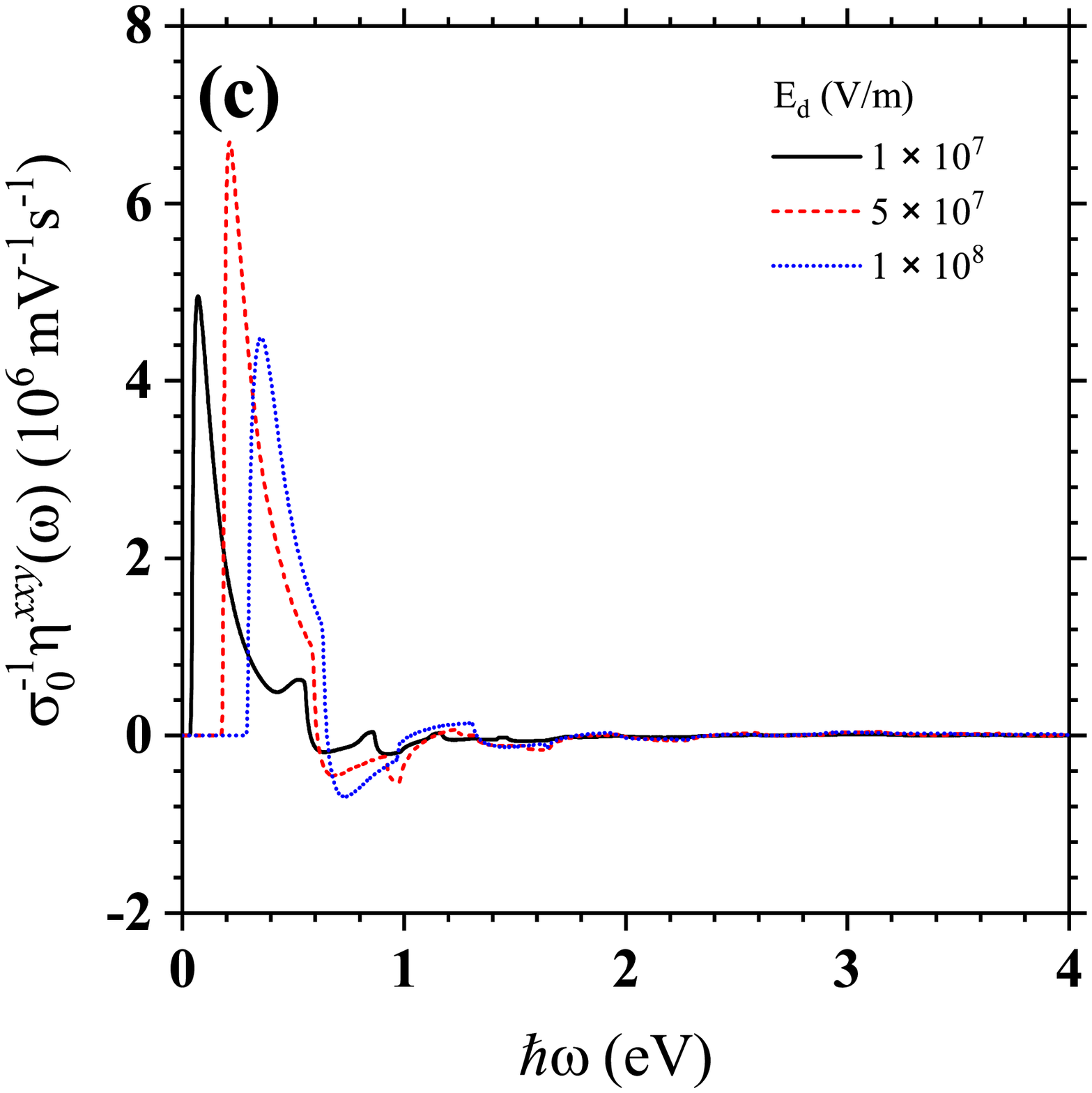}\includegraphics[width=7.5cm]{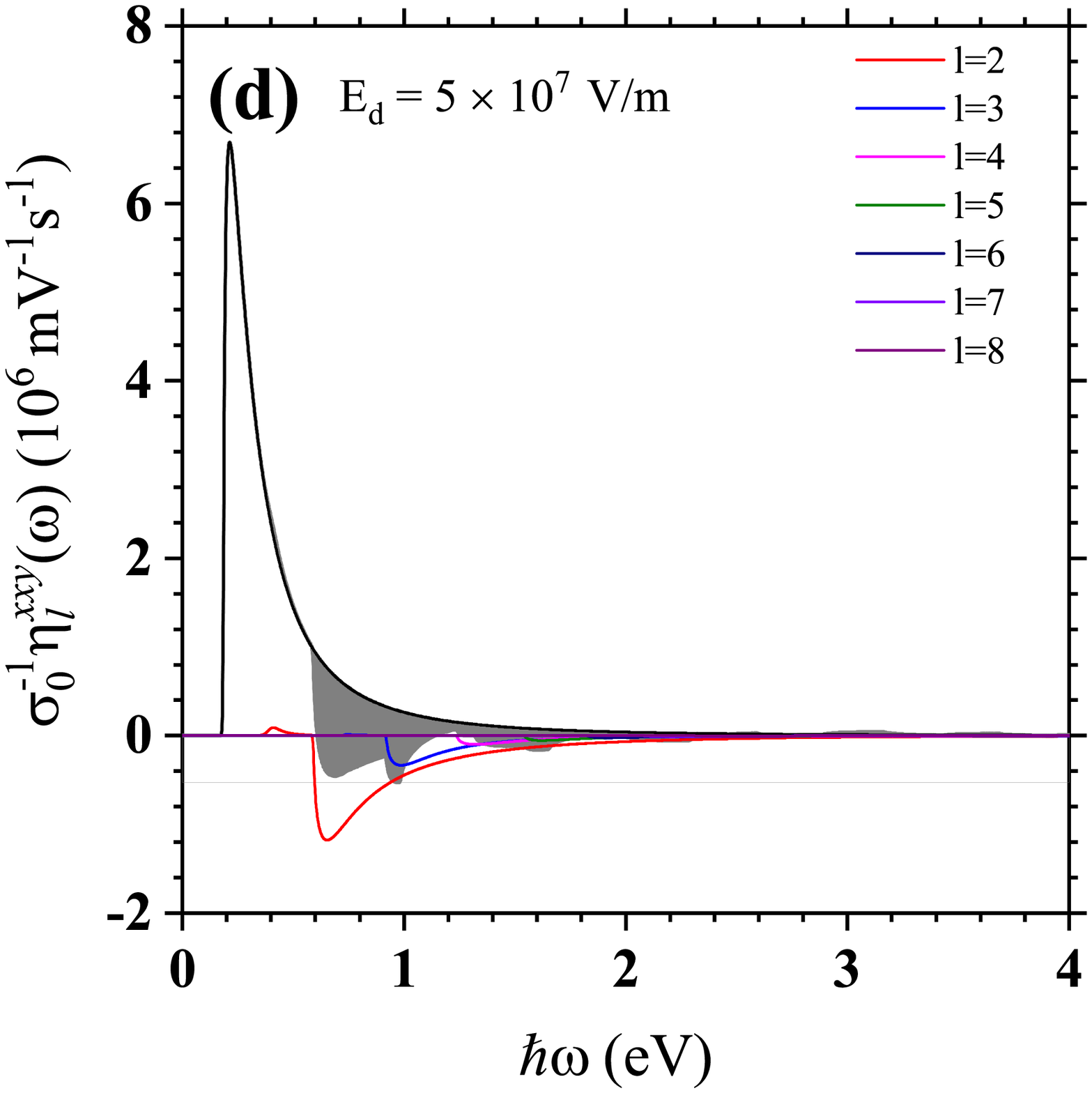}}
  \caption{(a) Spectra of injection coefficient $\tilde\eta^{xxyy}(\omega)$ for a $24$-zGNR for gates
    fields $E_d=10^2$, $10^4$, $10^6$,and 
    $10^7$~V/m at room temperature; the curves at the right of the
    vertical dashed line are scaled by 10 times. (b) The spectra of
    $\tilde\eta_{(+1)(-1)}^{xxyy}$ and $\tilde\eta_l^{xxyy}$ for $l=2,3,\cdots,8$
    for $E_d=10^2$ V/m. Two specific labels mark the separated contribution of transition from band $\pm 1$ to the band l = 3. (c) The injection coefficient $\eta^{xxy}$ of zGNR under large electric field  $E_d=10^7$, $5\times10^7$, and 
    $10^8$~V/m at room temperature. (d) The spectra of $\tilde\eta_l^{xxy}$ for $l=2,3,\cdots,8$
    for $E_d=5\times10^7$ V/m }
  \label{fig:eta} 
\end{figure*}
In Fig.~\ref{fig:jdos}(a) the energy differences
$\hbar\omega_{ss^\prime k}$ are
plotted with respect to $k$ for different band pairs $(s,s^\prime)$ with the
condition that $|f_{sk}-f_{s^\prime k}|\sim 1$.  The energy differences
$\hbar\omega_{s(-1)k}$ and $\hbar\omega_{s(-2)k}$ show valleys around
$k\sim g/3$ for all $s>1$, while
$\hbar\omega_{s(+1)k}$ shows valleys only for bands with $s\ge6$. These valleys determine the transition edge between these bands
and lead to divergent joint density of states from Eq.~(\ref{eq:deltaexact}). However, there is no such point for
$\hbar\omega_{(+1)(-1)k}$ at zero gate field. For nonzero gate field,
$\hbar\omega_{(+1)(-1)k}$ shows a valley at around similar k value $\sim g/3$, as discussed above. In Fig.~\ref{fig:jdos}(b), the gaps between
these band pairs are plotted as functions of the gate field, and the
color bar shows the $k$ values of the gap. The gate field modifies the gap
between the bands $(\pm1)$ significantly. 

Figure~\ref{fig:eta} gives the spectra of injection coefficients of a
$24$-zGNR at different $E_d$. In
general, the effects 
of a small $E_d$ can be treated perturbatively and the injection
coefficients can be connected with a third order sheet response
coefficients as 
\begin{align}
  \eta_{s_1s_2}^{xbc}(\omega) &= \tilde \eta_{s_1s_2}^{xbcy}(\omega)
                                E_d\,.\label{eq:etaW}
\end{align}
Figure~\ref{fig:eta} (a) plots
the spectra of $\tilde\eta^{xbcy}(\omega)$ for $E_d=10^2,
10^4, 10^6$ and $10^7$~V/m. When the photon energy is higher than the gap, the injection occurs. As the photon energy increases, the injection coefficient increases rapidly to the 
first peak, and afterwards it shows more peaks and the magnitude of each peak decreases with the photon energy. The first 5 peaks are located at around $\hbar\omega\sim 0.04$,
$0.53$, $0.85$, $1.16$, and $1.45$~eV; they slightly depend on the
broadening parameter $\Delta$ because the Dirac function is
approximated by a Gaussian function. When the photon energy is higher
than $2.5$~eV, the injection coefficients are about zero. As the
 field $E_d$ increases from $10^2$~V/m to $10^7$~V/m, the
value of $\tilde\eta^{xbxy}(\omega)$ changes little for
photon energies in certain windows ($\hbar\omega\in[0.2,0.4]$~eV and
a small energy range around 0.5~eV). Such energy window is enlarged to
$[0.1,0.6]$~eV if the gate field $E_d$ is between
$10^2$~V/m and $10^6$~V/m. The existence of these windows identifies the photon energies 
that the pertrubative treatment in
Eq.~(\ref{eq:etaW}) is appropriate. However, for photon energies
$\hbar\omega>1$~eV, although the injection coefficients are small, they differ significantly even for $E_d=10^2$~V/m and $10^4$~V/m,
indicating a non-perturbative feature of zGNR under electric fields. 

The peaks are mostly induced by the optical transitions associated with
the edge bands $s=\pm 1$, as shown in Fig.~\ref{fig:eta}(b), where the spectra of
$\tilde\eta^{xxyy}_{(+1)(-1)}(\omega)$ and
$\tilde\eta_l=\tilde\eta_{l(+1)}^{xxyy}+\tilde\eta_{l(-1)}^{xxyy}+\tilde\eta_{(+1)(-l)}^{xxyy}
+ \tilde\eta_{(+1)(-l)}^{xxyy}$
  are plotted for $E_d=10^4$~V/m. The electron-hole symmetry ensures  $\tilde \eta_{s_1s_2}^{xxyy}=\tilde \eta_{-s_2-s_1}^{xxyy}$ for an undoped ribbon. To better understand
these nonperturbative features, from
Eq.~(\ref{eq:eta}), we write the injection coefficient as  
\begin{align}
  \tilde \eta_{s_1s_2}^{xxyy}(\omega) = \frac{e^3}{E_dW\hbar^2}\sum_{j}
  \text{sgn}(\Delta^x_{s_{1}s_{2}k_j})\text{Im}[r^{y}_{s_{2}s_{1}k_j}r^{x}_{s_{1}s_{2}k_j}]f_{s_{2}s_{1}k_j}\,,\label {eq:eta1}
\end{align}
where $k_j$ are solutions of $\omega_{s_{1}s_{2}k_j}-\omega=0$ and
$\text{sgn}(x)$ is a sign function. In Eq.~(\ref{eq:eta1}) shows that the
joint density of states are cancelled out with the carrier velocity. For the contribution from the transitions between the
$s$th edge band and other bands $l\neq \pm 1$, the coefficients can be
obtained using the results in Appendix~\ref{app:berry} as
\begin{align}
  \tilde\eta_{ls}^{xxyy}(\omega)
  &= \frac{e^3}{E_dW\hbar^2} \frac{s}{2}\sum_j
    \sqrt{1-N_{k_j}^2}\text{Im}[\xi_{s^\prime lk_j}^{0;x}\xi_{(-s^\prime)lk_j}^{0;y}](-f_{sk_j})
\end{align}

As an example, the spectra of
$\tilde \eta_{(+3)(+1)}$ and $\tilde \eta_{(+3)(-1)}$ are shown in
Fig.~\ref{fig:eta} (b). Their values are nearly opposite thus their sum is much smaller, which indicates an interesting cancellation
between the transitions. Because the nearly degenerate edge bands, the
dependence on $E_d$ of the injection coefficients is complicated.

For higher gate fields, the band structures are dramatically changed,
and the understanding of the current injection cannot be based on the
quantities of ungated ribbons. The contribution from $\tilde \eta_{s(+1)}$ becomes
negligible because there is less occupuation on the band $s=1$.  Figures~\ref{fig:eta}(c,d) give the spectra of
$\eta^{xxy}(\omega)$ at $E_d=10^7$, $5\times 10^7$, and
$10^8$~V/m. For low photon energy, the injection occurs between the
bands $s=-1$ and $s=1$. As the electric field increases from $10^7$ to
$10^8$~V/m, the injection coefficients keep almost unchanged,
instead, the peak position changes significantly, indicating the
changes of the band structure. Similar to the cases at small gate
fields, the injection coefficients decrease with the
photon energy quickly.

We give an estimation on how large the injection current can be at a gate field $10^6$ V/m. At the photon energy 0.55 eV around the second peak, our calculated current injection rate is about 0.1~m$^2$V$^{-2}$s$^{-1}$, it corresponds to the bulk current injection rate $\sim 2\times 10^{10}$ $\mu$As$^{-1}$V$^{-2}$  considering the 0.3 nm thickness of zGNR, which is nearly 25 times larger than that in bulk GaAs\cite{PhysRevB.74.035201}. In this case, a laser pulse with intensity 0.1 GW/cm$^2$ and duration 1 ps can generate an injection current $\sim$1.1 $\mu$A. 

\subsection{Shift conductivity of $24$-zGNR}
\begin{widetext}
\begin{figure*}[htp]
  \centering
  \includegraphics[width=5.5cm]{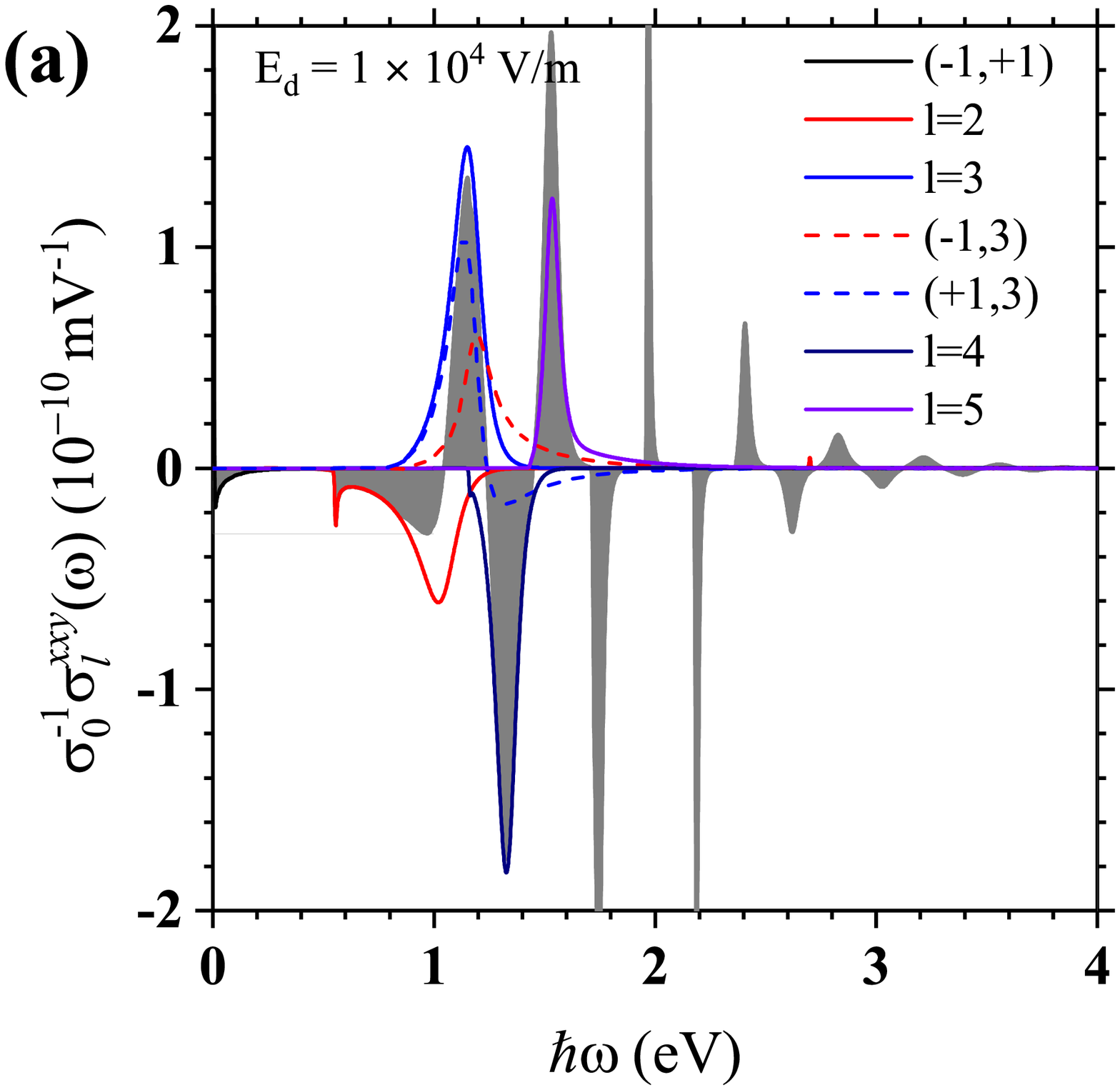}\includegraphics[width=5.5cm]{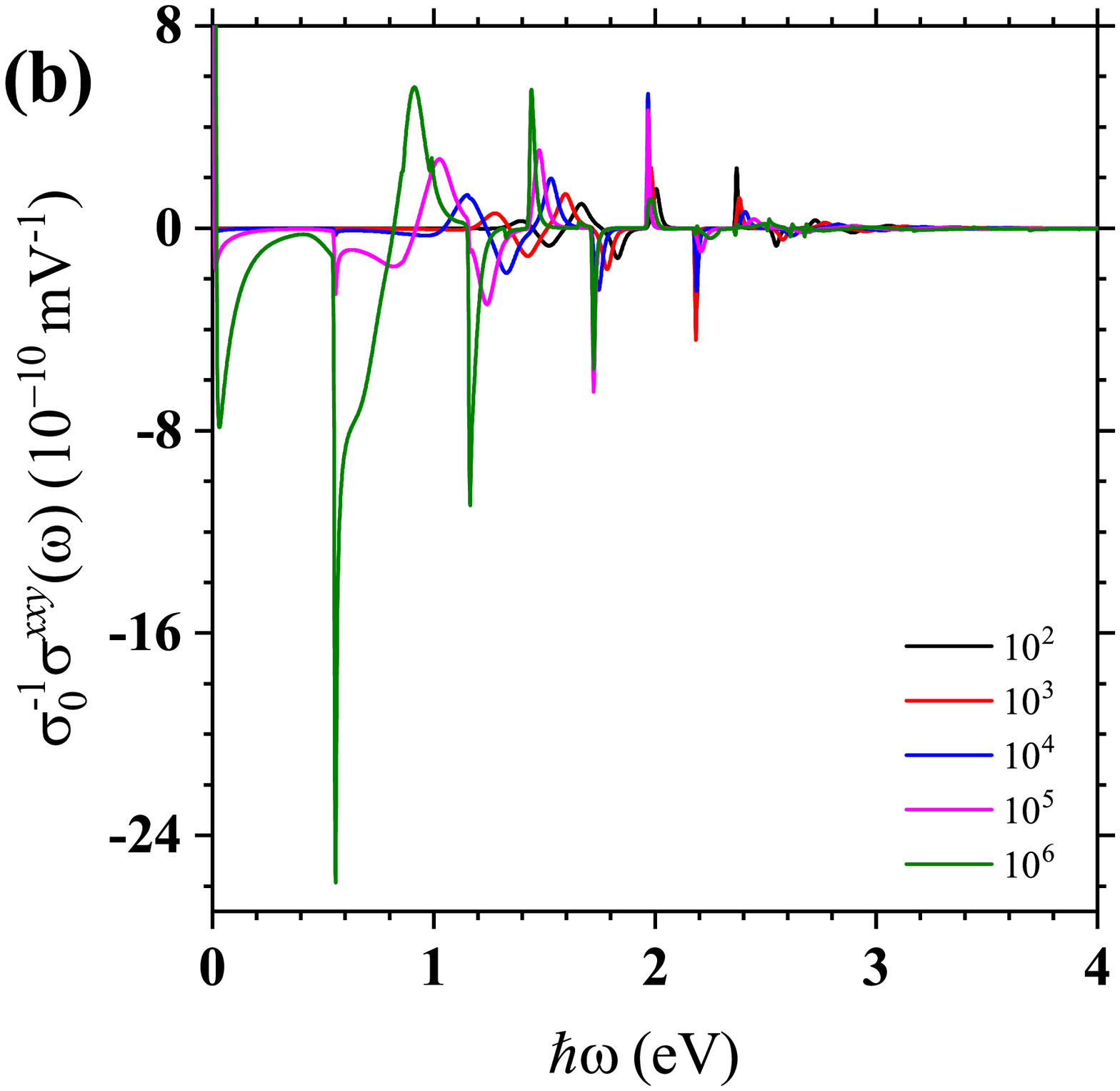}\includegraphics[width=5.5cm]{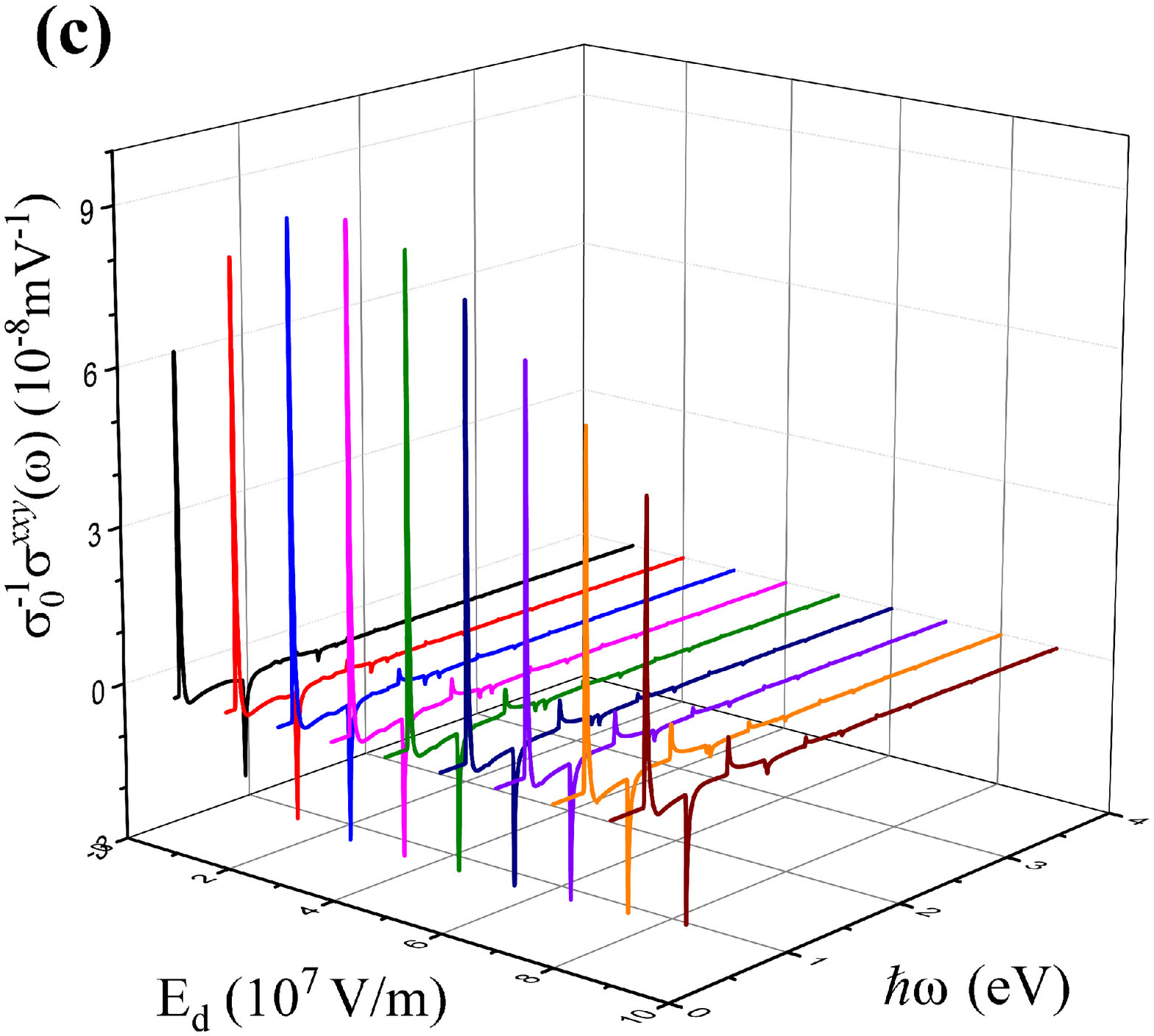}
  \caption{Spectra of shift conductivity $\sigma^{xxy}(\omega)$ for
    an undoped $24$-zGNR at different gate fields. (a) Transition resolved contribution of $\sigma^{xxy}(\omega)$
     at $E_d=10^4$~V/m. The shadowed region gives the total conductivity. The plotted contribution from different band pairs are $ \sigma^{xxy}_{(+1)(-1)}(\omega)$,
     $\sigma^{xxy}_{(+3)(\pm1)}(\omega)$, as well as
       ${\sigma}^{xxy}_{l}(\omega) =
       \sum_{\pm}{\sigma}^{xxy}_{(+l)(\pm 1)}(\omega) +
       {\sigma}^{xxy}_{(\pm1)(-l)}(\omega)$ for
       $l=2,3,4,5$. (b) Spectra of $\sigma^{xxy}(\omega)$ at $E_d=10^2$,
     $10^3$, $10^4$, $10^5$, and  
     $10^6$~V/m at room temperature. (c)  Spectra of
     $\sigma^{xxy}(\omega)$ at gate fields up to $10^8$~V/m.}
  \label{fig.:sig} 
\end{figure*}
\end{widetext}

Figure~\ref{fig.:sig} (a) gives spectra of
$\sigma^{xxy}(\omega)$ as well as the contributions from different
optical transitions for a gate field $E_d=10^4$~V/m. The spectra
show the following features: (1) The values of the shift conductivity
decrease quickly with the photon energy for
$\hbar\omega<0.5$~eV, and drop suddenly at
$\hbar\omega\sim0.55$~eV to a very sharp valley, which is induced by
the divergent joint density of states between the bands $\pm 1$ and $\pm2$. (2) With increasing the photon
energy, the conductivity shows positive peaks and negative
valleys alternatively. The first four valleys locate at $0.97$, $1.33$,
$1.74$, and $2.19$~eV, and the first four peaks locate at $1.15$,
$1.53$, $1.97$, and $2.41$~eV; other peaks and valleys have much smaller
amplitudes. (3) The peaks and valleys have different widths, and the widths for the third
peak and the fourth valley are very narrow. These peaks and valleys can
be better understood from transition resolved conductivities, which
are also plotted in Fig.~\ref{fig.:sig} (a)  for  $\sigma^{xxy}_{(+1)(-1)}(\omega)$, $\sigma^{xxy}_{(+3)(\pm1)}(\omega)$, as well as
       $ {\sigma}^{xxy}_{l}(\omega) =
       \sum_{\pm} {\sigma}^{xxy}_{(+l)(\pm 1)}(\omega) +
        {\sigma}^{xxy}_{(\pm1)(-l)}(\omega)$ for
       $l=2,3,4,5$. Similar to the injection processes,
$\sigma_{(+s)(+s^\prime)}^{xxy}(\omega)
=\sigma_{(-s^\prime)(-s)}^{xxy}(\omega)$ holds for an undoped ribbon. However, different
from the injection process, the values of $\sigma_{(+s)(\pm1)}^{xxy}$ and
$\sigma_{(\pm1)(-s)}^{xxy}$ have similar amplitudes and same signs
but locate at different photon energies, and their total contribution leads to a wider peak
or valley comparing those in the injection coefficients shown in
Fig.~\ref{fig:eta} (b). 
The transition $\sigma^{xxy}_{2}$ is composed of two
valleys: one is at lower photon energy, which is induced by the divergent
joint density of states at 0.56~eV, and
the other is at higher photon energy around 1~eV.

In Fig.~\ref{fig.:sig} (b) the shift conductivities for $E_d=10^2$,
$10^3$, $10^5$, and $10^6$~V/m are  plotted for a comparison. Similar to the
injection processes, the shift
conductivities for photon energies lower than
0.6~eV are mostly contributed from the transition between two
edge bands, and they are linearly proportional to the gate field. 
As the gate field $E_d$ increases from $10^2$~V/m to $10^6$~V/m, the
location of the first valley does not change because of the negligible
bandgap shift, but the peak value increases linearly from
$2.5\times10^{-12}$~m$^2$/V$^2$ to $2.5\times10^{-8}$~m$^2$/V$^2$. 
For photon
energies higher than 0.6~eV, despite of 4 orders
of magnitude change for the gate field, the values for the shift conductivity are almost at the
same order of magnitude; this indicates a nonperturbative dependence
on the gate field, which is again induced by the near degeneracy of the edge
states. Besides, the locations of peaks and valleys shift to lower
photon energies as the gate field increases.  
Figure~\ref{fig.:sig} (c) gives the spectra of the shift conductivity
for gate field up to $1\times 10^8$~V/m. For large $E_d$, 
the values around the first two peaks are much larger; the first peak
value shows a maximum around $E_d = 4\times 10^7$~V/m, while the value of the second valley changes little.

As the case of injection current, we estimate the magnitude of the shift current of zGNR for a gate field $10^6$ V/m. At the photon energy 0.56 eV around one of the valleys, the sheet shift conductivity is $1.57 \times 10^{-13}$  AmV$^{-2}$. It corresponds to the bulk photocurrent conductivity $524 \ \mu$AV$^{-2}$, which is twice larger than that in 2D GeSe ($200 \ \mu$AV$^{-2}$). \cite{PhysRevLett.119.067402}
A laser intensity 0.1 GW/cm$^2$ can generate  a shift current
$\sim$0.29 $\mu$A, a few times smaller than injection currents.
\begin{widetext}
\begin{figure*}[!ht]\centering
\includegraphics[width=7.5cm]{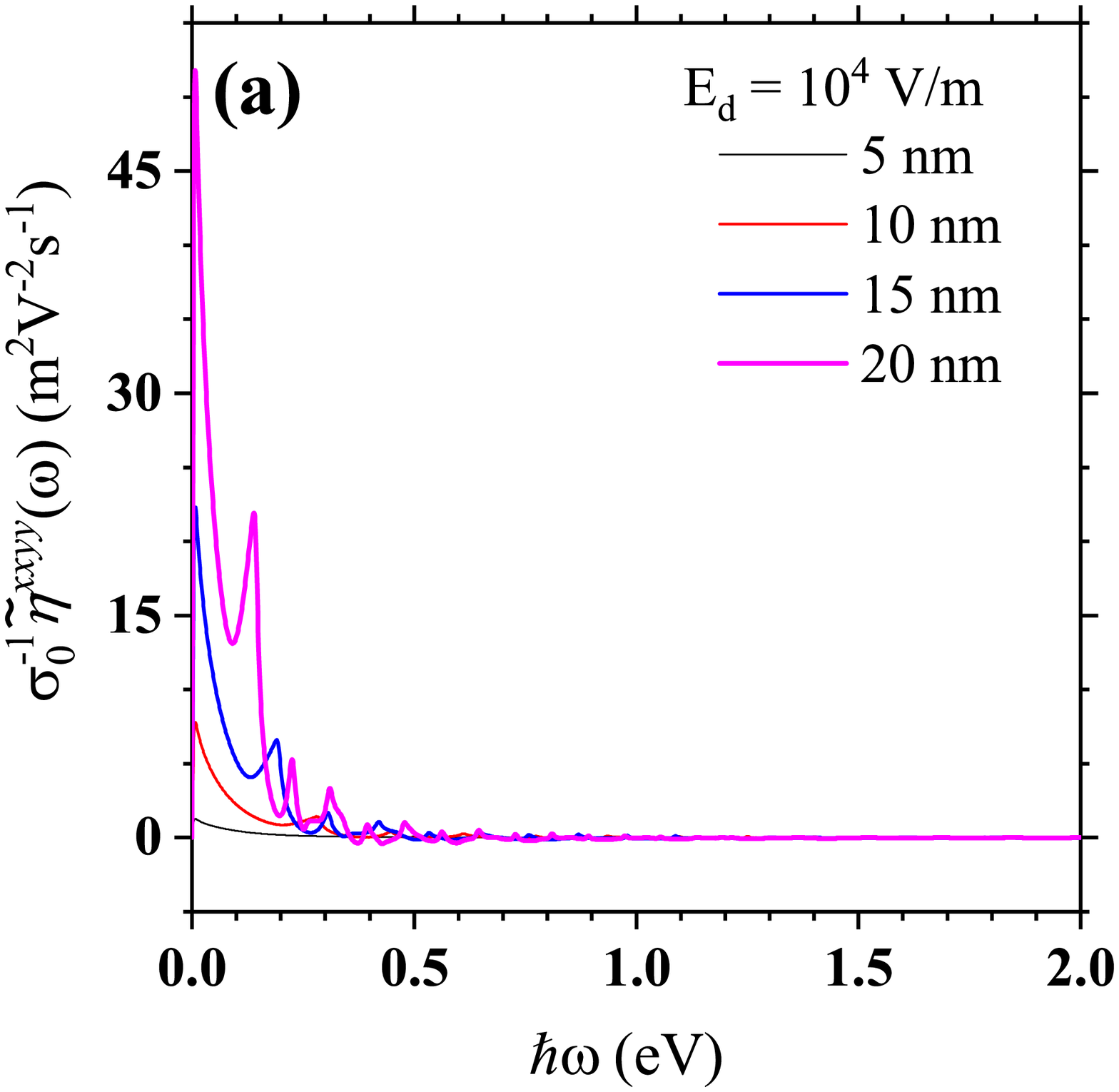}\includegraphics[width=7.5cm]{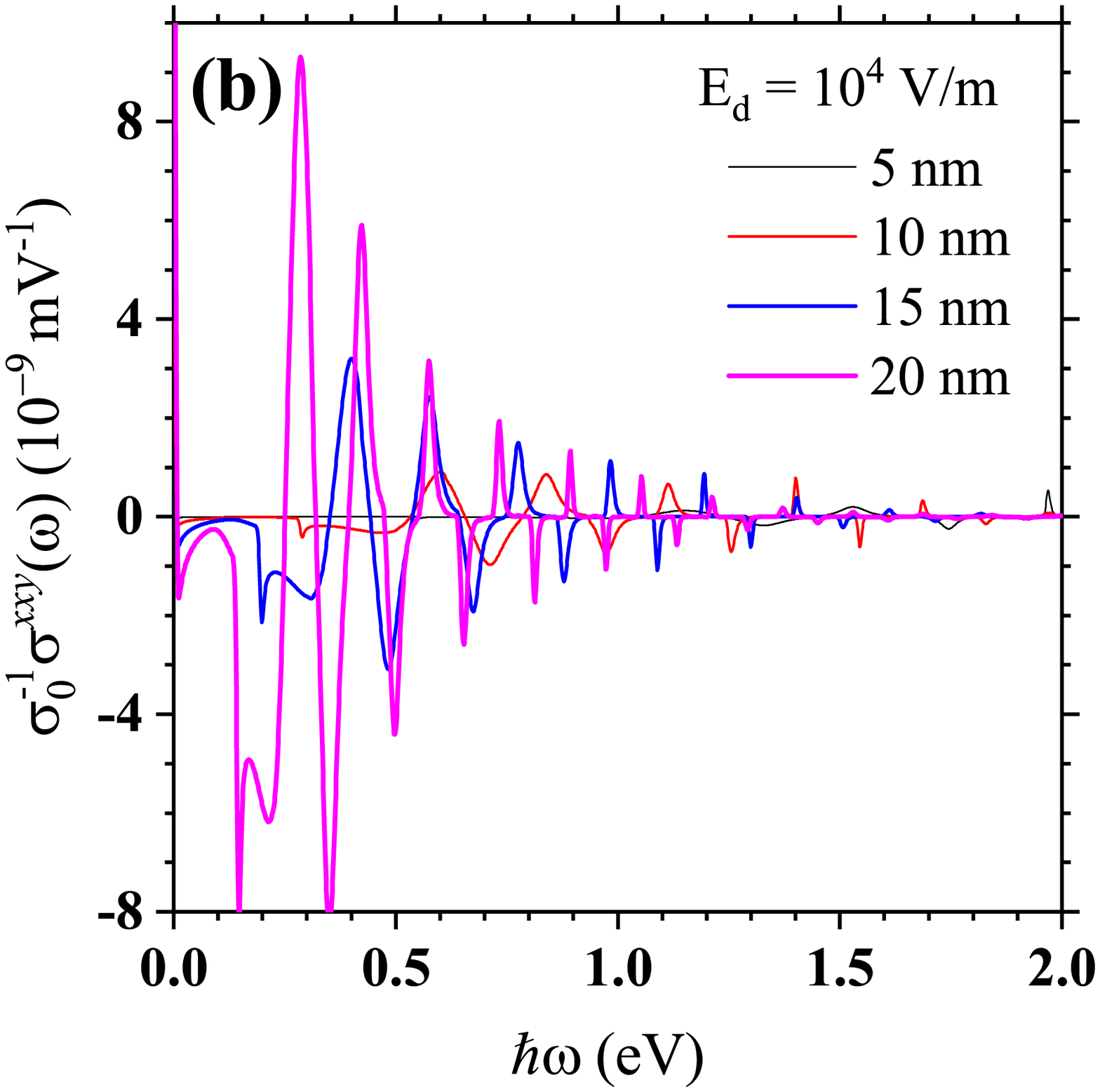}\\
\includegraphics[width=7.5cm]{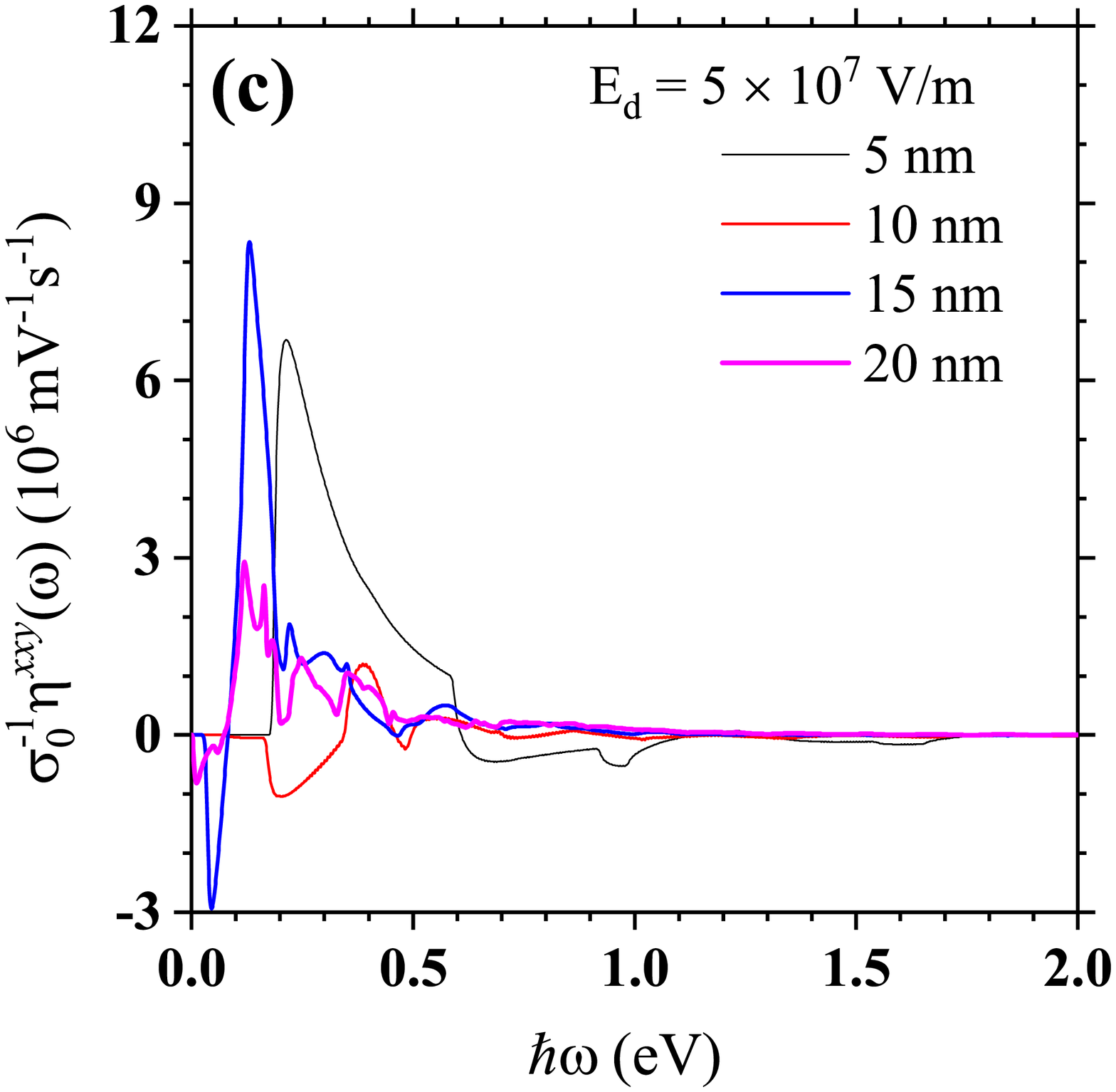}\includegraphics[width=7.5cm]{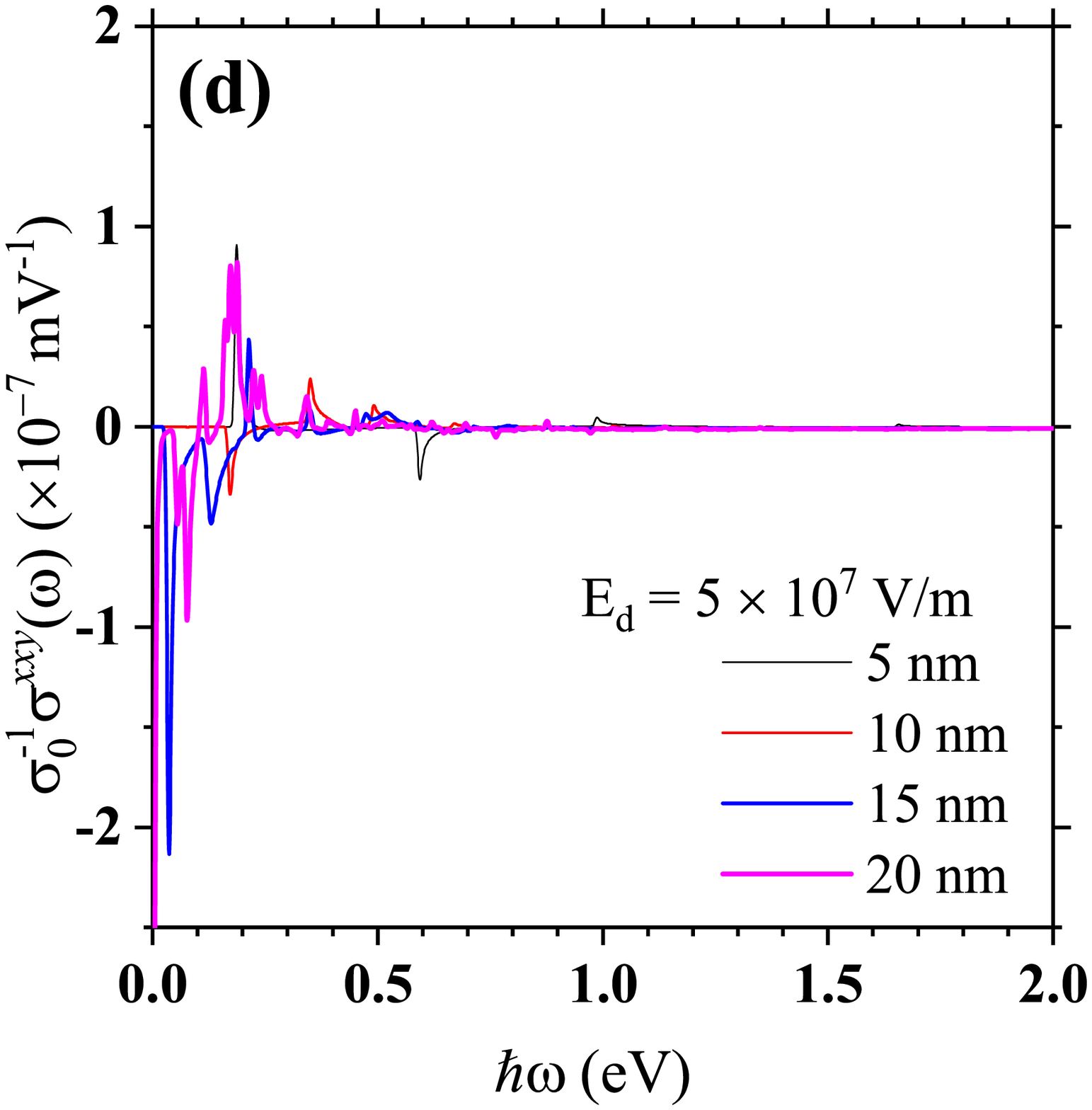}
\caption{The spectra of injection coefficients and shift
  conductivities for different ribbon width $W=5$, $10$, $15$, and
  $20$~nm. (a) $\tilde{\eta}^{xxy}(\omega)$ at $E_d=10^4$~V/m, (b)
  $\sigma^{xxy}(\omega)$ at $E_d=10^4$~V/m, (c)
  $\tilde{\eta}^{xxy}(\omega)$ at $E_d=5\times 10^7$~V/m, (b) 
  $\sigma^{xxy}(\omega)$ at $E_d=5\times 10^7$~V/m.}
\label{fig:wd}
\end{figure*}
\end{widetext}

\section{Width dependence\label{sec:width}}
Figure~\ref{fig:wd} gives injection coefficients and shift
conductivities for zGNR with different widths $W =$ 5,10,15,and 20 nm
(corresponding to $N=$24,48,72 and 96) at two
gate fields $E_d=10^4$~V/m and $5\times 10^7$~V/m. With the increase
of the ribbon width, there appear more subbands, and the energy difference of neighbour bands
decreases. Therefore, both for the injection coefficients and for shift conductivities, there exist more peaks or valleys in the
spectra with the increase of the ribbon width, while their amplitudes change little. A wider ribbon can generate
larger currents. 

\section{Conclusion}\label{sec:conclusion}
Based on a simple tight binding model, we explored
the one-color injection currents and shift currents in zigzag graphene nanoribbons, where a gate field across the ribbon is applied to break the
inversion symmetry. The gate field lifts the degeneracy of
the edge bands and significantly modifies their wave functions, which
leads to the nonperturbative behavior with respect to even very weak gate field. The spectra of injection coefficients and shift
conductivities show fruitful structures, including many peaks and
valleys, with locations strongly depending on the
ribbon width. These fine structures indicate the importance of
the contributions from different bands. The injection coefficients are
almost positive for different photon energies, while the sign of the
shift conductivities is very sensitive on the photon energies. 
Under excitation by a pulsed laser with intensity 0.1 GW/cm$^2$, our calculation for a 5 nm wide zGNR shows that the injection current reaches $\sim$1.1 $\mu$A for a pulse with duration 1 ps, whereas the shift current is $\sim$0.29 $\mu$A. Because the injection current and the shift current can be separately
excited using light with different polarization, and their magnitudes can be
well tuned by the static electric field strength, these features
could be experimentally observed. 

\acknowledgements
This work has been supported by Scientific research project of
the Chinese Academy of Sciences Grant No. QYZDB-SSW-SYS038, National
Natural Science Foundation of China Grant
No. 11774340, 11974093 and 12034003. J.L.C. acknowledges the support from ``Xu
Guang'' Talent Program of CIOMP. Y.D.W. thanks Kaijuan Pang for the help on diagrams.

\appendix

\section{Berry connections of edge states\label{app:berry}}
When there is no gate field, the wave functions can be chosen to
satisfy
\begin{align}
  \xi_{(+1)(-1)k}^{0;x} &=0\,,\\
  \xi_{(+1)(-1)k}^{0;y} &=\xi_{(-1)(+1)k}^{0;y} \text{ as real numbers}\,.
\end{align}
From Fig.~\ref{fig:bands} we have calculated the results of the left hand side of
\begin{align}
  {\cal R}_{smk}^{0;cx} -\partial_k r_{smk}^{0;c} = i
  (\xi_{ssk}^{0;x}-\xi_{mmk}^{0;x})r_{smk}^{0;c}\,,\text{ for } m=\pm 1,\pm2,\cdots\,.
\end{align}
and found that all of them are zero in our numerical
resolution. Thus in the following we will adopt
$\xi_{ssk}^{0;x}-\xi_{llk}^{0;x}=0$ without giving an exact derivation. In the
new basis of $\{|sk\rangle,|lk\rangle^0\}$, the position matrix
elements are 
\begin{align}
  \langle sk|\tilde{\bm r}_k + i\hat{\bm x}\partial_k |lk\rangle^0
  &=\frac{1}{\sqrt{2}}\left[s\sqrt{1+sN_k}\bm\xi_{(+1)lk}^0+\sqrt{1-sN_k}\bm\xi_{(-1)lk}^0\right]\,.
\end{align} 
With the inclusion of the gate field,  the diagonal Berry connections can be written as
\begin{align}
\xi^x_{ssk}=&  \langle sk|\tilde{r}^x_k + i\partial_k
   |sk\rangle
    = \frac{i}{2}\left[(1+sN_k)\xi_{(+1)(+1)k}^{0;x}+(1-sN_k)\xi^{0;x}_{(-1)(-1)k}\right]\,,
\end{align}
then we get
\begin{align}
  \xi^x_{(+1)(+1)k}-\xi^x_{(-1)(-1)k} &=i N_k\left[\xi^{0;x}_{(+1)(+1)k}-\xi^{0;x}_{(-1)(-1)k}\right]=0\,.
\end{align}
The off-diagonal Berry connections are
\begin{align}
  \xi^x_{(+1)(-1)k} 
  =&\frac{i}{2}\frac{\partial_kN_k}{\sqrt{1-N_k^2}}\,,\\
     \xi_{(+1)(-1)k}^y
  =&N_k\xi_{(+1)(-1)k}^{0;y}\,.     
\end{align}
Further we can calculate
\begin{align}
  {\cal R}_{(+1)(-1)k}^{cx} &= \partial_k \xi^c_{(+1)(-1)k}\,,\\
  {\cal R}_{slk}^{cx}&=\partial_k \xi^c_{slk}\,.
\end{align}

\bibliographystyle{apsrev4-1}
\bibliography{article}

\end{document}